\documentclass{LMCS}
\usepackage{graphicx}
\usepackage{latexsym}
\usepackage{amsfonts}
\usepackage{amsmath}
\usepackage[all]{xy}
\usepackage{array}
\usepackage{url}  
\usepackage{epic}

\usepackage{enumerate,hyperref}

\newif\ifLMCS\LMCStrue
\newif\iflncs\lncsfalse
\usepackage{amsthm}

\newcommand{\com}{\newcommand}

\newcommand{\val}{\mathchoice{\mbox{\it Val}}{\mbox{\it Val}}%
{\mbox{\scriptsize\it Val}}{\mbox{\scriptsize\it Val}}}
\renewcommand{\vec}[1]{\mathbf{#1}}

\newcommand{\bell}{\mbox{Bell}}
\newcommand{\trans}[3]{{#2 \stackrel{#1}{\mapsto} #3}}
\newcommand{\edge}{\leftrightarrow }
\newcommand{\strict}{\xrightarrow{{}^{>}}}
\newcommand{\nonstrict}{\xrightarrow{{}^{\ge}}}
\com{\ints}{{\mathbb Z}}
\com{\nats}{{\mathbb N}}

\let\cal=\mathcal

\com{\pgt}[1]{{\tt #1}}
\com\powerset{{\cal P}}
\com\cv{{\circ}}  

\ifLMCS
\theoremstyle{plain}\newtheorem{claim}[thm]{Claim}
\else
 \newtheorem{thm}{THEOREM}[section]
 \newtheorem{obs}[thm]{OBSERVATION}
 
 \newtheorem{lem}[thm]{LEMMA}
 
 \theoremstyle{defi}
 \newtheorem{defi}[thm]{Definition}
 
 \newtheorem{algo}{Algorithm}[section]
 \newtheorem{exa}{Example}[section]
\fi
\com{\bthm}{\begin{thm}}
\com{\ethm}{\end{thm}}
\com{\bdfn}{\begin{defi}}
\com{\edfn}{\end{defi}}
\com{\blem}{\begin{lem}}
\com{\elem}{\end{lem}}
\com{\bex}{\begin{exa}\pushQED{\qed}}
\com{\eex}{\popQED\end{exa}}
\com{\bprf}{\begin{proof}}
\com{\eprf}{\end{proof}}

\newcommand{\bi}{\begin{itemize}}
\newcommand{\ei}{\end{itemize}}
\newcommand{\be}{\begin{enumerate}}
\newcommand{\ee}{\end{enumerate}}

\com{\fl}{\noindent}
\com{\hair}{\hspace{3mm}}
\com{\vair}{\vspace{3mm}}

\newenvironment{fig0}[3]
{
\xdef\fighack{\noexpand\caption{#1}\noexpand\label{#3}}
\begin{figure*}[#2]
\hspace*{3mm}\begin{minipage}{0.95\textwidth}}{\end{minipage}%
\fighack%
\end{figure*}}

\title{Size-Change Termination, Monotonicity Constraints and Ranking Functions}

\ifLMCS
 \author[Ben-Amram]{Amir M. Ben-Amram}
 \address{School of Computer Science, The Tel-Aviv Academic College, Israel}	
 \email{benamram.amir@gmail.com}  

\fi

\def\doi{6 (3:2) 2010}
\lmcsheading%
{\doi}
{1--32}
{}
{}
{Sep.~29, 2009}
{Sep.~21, 2010}
{}   

\begin{document}

\maketitle

\begin{revision}
  This is a revised and corrected version of the article originally
  published on July 11, 2010.
\end{revision}

\begin{abstract}
%
Size-Change Termination (SCT) is a method of proving
program termination based on the impossibility 
of infinite descent. To this end we may use a program abstraction in which
transitions are described by
\emph{monotonicity constraints} over (abstract) variables.
Size-change graphs are a subclass where
 only constraints of the form $x>y'$ and $x\ge y'$ are allowed.
 Both theory and practice are now more
evolved in this restricted framework than in the general framework
of  monotonicity constraints. 
This paper shows that it is possible to adapt and extend some theory from
the domain of size-change graphs to the general case, thus complementing
previous work on monotonicity constraints.
In particular, we present precise decision procedures for termination;
and we provide a procedure to construct explicit
global ranking functions from monotonicity constraints
in singly-exponential time, which is better than
what has been published so far even for size-change graphs.
\end{abstract}
\ifLMCS\else
ACM subject classes: D.2.4; F.3.1
\fi

\section{Introduction}
\label{sec:intro}

This paper is concerned with
termination analysis. This is a fundamental and much-studied problem of software
verification, certification and transformation. 
While being, historically, the epitome of undecidability,
much progress has been made on automatizing termination proofs, so that now we
find termination assertions in practical programming tools such as
JML~\cite{Leavens-Baker-Ruby99b}
and proof assistants such as ACL2~\cite{MV-cav06} and
Isabelle~\cite{KraussHeller09}.
One of the contributing factors to
the development of automated termination analysis was 
its importance 
in designing certain meta-programs,
e.g., interpreters for Logic Programs~\cite{Naish1985,DD:94}
and partial evaluators~\cite{Jo:88,JGS:93}.
In such applications, the danger is that the meta-program will diverge,
which is unacceptable. In theorem provers, termination proofs are required
for showing that recursive definitions are sound.

A subproblem of termination analysis is the construction of
\emph{global ranking functions}. Such a function is required to decrease
in each step of a program (for ``step" read
basic block, function call, etc, as appropriate); 
the function witnesses
the progress towards termination.

An explicitly presented ranking
function whose descent is (relatively) easy to verify
is a useful \emph{certificate} for termination~\cite{krauss07} and may have other uses,
such as running-time analysis~\cite{AAGP:sas08}.

As the general halting problem is undecidable, every method for termination
analysis consists, in principle, of identifying some {\em subproblem\/}
that is decidable. This can be done in a more or less structured manner;
a very structured approach is to break the termination problem for
programs into two stages: the first is 
\emph{abstraction} of the program, in which the concrete
program is replaced with an abstract one, essentially a simplified model of the
original. The second stage is \emph{analysis}
of the abstract program. One benefit of this separation is that the abstract
programs may be rather independent of the concrete programming language.
Another one is that the termination problem for the abstract programs may be
decidable. 

Size-change termination (SCT~\cite{leejonesbenamram01}) is such an approach.
It views a program as a transition system with states. The abstraction
consists in forming a \emph{control-flow graph}
for the program, identifying a set of \emph{state variables},  and forming a finite set of 
\emph{size-change graphs} that are abstractions of the transitions of the program.
In essence, a size-change graph is a set of inequalities between variables
of the source state and the target state.
Thus, the SCT abstraction is an example of a transition system
defined by \emph{constraints} of a particular type.

The technique concentrates on 
well-founded domains, where infinite descent is impossible.
An SCT termination proof is a proof that any (hypothetical) infinite run would
decrease some value monotonically and endlessly so that well-foundedness is
contradicted.
Since only chains of non-increasing values are of interest,
only two types of inequalities were admitted into the constraints
in~\cite{leejonesbenamram01}: $x>y'$ (old value of $x$ greater than new value
of $y$) and $x\ge y'$.

Size-change graphs lend themselves to a very natural generalization: 
Monotonicity Constraints. Here, a transition may be described by any
conjunction of order relations, including
equalities as well as strict and non-strict inqualities, and involving 
any pair of variables from the source state and target state.  Thus,
it can express
a relation among source variables, that applies to states
in which the transition may be taken; a relation among the target variables,
which applies to states which the transition may produce;
and, as in SCT, relations involving a source
variable and a target variable, but here
equalities can be used, as well as relations like $x\le x'$, that is, an
increase.

The \emph{Monotonicity Constraint Systems} treated in this paper will include
another convenience, \emph{state invariants} associated with a point in the
control-flow graph. These too
are conjunctions of order constraints.

Monotonicity constraint systems generalize the SCT abstraction
and are clearly more expressive. It may happen that
analysis of a program yields
monotonicity constraints
which are not size-change graphs; in such a case, simply approximating the constraints by a size-change
graph may end up missing the termination proof. For an example, see the next section.
It is not surprising, perhaps, that
Monotonicity Constraints actually predated the SCT
framework---consider the Prolog termination analyses in~\cite{LS:97,CT:99}.
But as often happens in science, concentrating on a simplified system that
is sufficiently interesting was conducive to research,
and the formulation of the SCT framework
led to a series of interesting discoveries.
To pick up some of the salient points:

\begin{enumerate}[$\bullet$]
\item
 The SCT abstraction has a simple semantics, in terms of transition systems.
\item
 It has a useful combinatorial representation as a set of graphs,
known as the size-change graphs.
\item
 A termination criterion has been formulated in terms of these graphs
(the existence of an infinite descending thread in every infinite multipath
\cite{leejonesbenamram01}).
\item
 This criterion is equivalent to the termination of every model (transition
system)---in logical terms, this condition is sound and complete
\cite{leejonesbenamram01,Lee:ranking}.
\item
 Termination of a set of size-change graphs can be effectively decided;
while having exponential worst-case time, the
method is often usable.  The best-known method is a
closure-based algorithm~\cite{leejonesbenamram01}, which can be theoretically related to
 termination proofs by  disjunctively well-founded transition 
invariants~\cite{PR-LICS04,Codish-et-al:05}. More theoretically, the
complexity of the problem has been identifed as PSPACE-complete
\cite{leejonesbenamram01}.
\item
 It has been established that a global ranking function can also be 
effectively constructed from the size-change graphs. Lee~\cite{Lee:ranking}
gave the first proof, where the size of the resulting ranking expression
is up to triply-exponential in the size of the
abstract program. This left open the challenging problem
of improving this upper bound.
 Progress regarding certain special cases
is reported in~\cite{BL:ranking2}.
\end{enumerate}

Which of the useful results can also be obtained in the stronger
framework of monotonicity constraints?  
One contribution of this paper is an answer:
\emph{in essence, all of them}.

	A second contribution of this paper is an algorithm to verify termination
of a monotonicity constraint system while constructing a global ranking
function, all in singly-exponential time. Thus, we solve the open problem
from~\cite{Lee:ranking}, and, surprisingly, by tackling a super-problem%
\footnote{The author is aware of another singly-exponential solution
for SCT, communicated privately by Chin Soon Lee.}.
The ranking functions generated have a simple form, and are based solely on lexicographic
descent. Their worst-case size as well as ordinal height match known lower bounds.



\subsubsection*{Prior Work}

The Prolog termination analysers \emph{Termilog}~\cite{LS:97,LSS:04} and 
\emph{Terminweb}~\cite{CT:99}
can both construct monotonicity constraints for a program. The abstraction used in both systems
also includes instantiation patterns,
particular to logic programming, however ignoring
this aspect, the termination analysis boils down to deciding
termination of monotonicity constraint systems.

The implementation of SCT analysis in ACL2~\cite{MV-cav06} added to the original
SCT formulation a notion of state invariants (in the form of ``calling contexts").
This predates our inclusion of state invariants in the monotonicity-constraint abstraction.

Codish, Lagoon and Stuckey~\cite{Codish-et-al:05} were the first to investigate
the question of monotonicity constraints versus size-change graphs.
They made the 
intriguing observation, that the termination test used 
in~\cite{LS:97,CT:99,LSS:04} is
sound and complete for SCT, but incomplete for general monotonicity constraints.
They also presented a correct test (that is, sound and complete),
closely related to
the algorithms presented here (see Section~\ref{sec:stable}).

\subsubsection*{Usefulness of monotonicity constraints}

Another important point in~\cite{Codish-et-al:05} is how monotonicity
constraints can imply termination for the integer domain, in a way that transcends the
well-founded model discussed above.  For example, monotonicity constraints are
sufficient for deducing termination of a loop such as 
\verb/while(x<y) x=x+1/.  The theory presented in this paper can, in principle, be adapted
to the integer domain. The basic ideas can be found in~\cite{Codish-et-al:05} 
and~\cite{BA:cav09}. The details are, however, complicated, and in order to keep this
paper lucid (and of reasonable length),
 they are deferred to a future publication. Since this paper
is an attempt to present a coherent
theory for monotonicity constraints, including several novel definitions
and results, it will be restricted to
the well-founded model, which allows for a more elegant theoretical 
development.  It may be true that for proving termination in the cases
where the well-founded model is natural (programs operating over lists,
trees etc.), SCT works so well that the extension is practically redundant
(at least, I have not seen empirical evidence to the contrary). Nonetheless,
users of SCT in such contexts may benefit of the new ranking-function
construction, as no singly-exponential worst-case construction has been known
for SCT; and future application to the integer domain will be facilitated
by the theoretical foundation. 

The reader is asked to bear, therefore,
with the rather abstract presentation in this paper, and turn to~\cite{leejonesbenamram01} 
or other papers on SCT for concrete examples of programs and their abstract representation.
All examples proved to teminate using size-change graphs will also be amenable to the new
ranking-function construction.

Some of the results in this paper have been presented at CAV 2009~\cite{BA:cav09}.

\subsubsection*{How to abstract a program}
Let us conclude this discussion by saying something about the method of abstracting programs,
that is, extracting the monotonicity constraints. 
Such an analysis may be done in a very simple-minded manner
in the style of the toy example given in~\cite{leejonesbenamram01}, but when it is necessary
to establish relations between
\emph{computed values}, a stronger static analysis will be necessary. Fortunately, such tools
exist.
A classic example is~\cite{BrodskySagiv:91} for
logic programs, which underlies Termilog. The more recent~\cite{BenoyKing:96} is used in
Terminweb, as is polyhedral analysis~\cite{CousotHalbwachs:1978} which is an all-times favourite
for handling imperative programs~\cite{Avery:06,Albert.et.al-fmoods08,SpotoMP09}.

The implementations of SCT analysis in ACL2~\cite{MV-cav06} 
and Isabelle~\cite{krauss07}
rely on a theorem prover to derive constraints. 

In the rest of this paper, we concentrate on analysing the abstract programs, and do not aim
to contribute to the art of abstraction, except indirectly, by drawing attention to the possibilities
of abstracting a program to a monotonicity constraint system.

\section{Basic definitions and examples}

This section introduces monotonicity constraint systems (MCS) and
their semantics, and formally relates them to the SCT abstraction.
As terminology is not uniform
across SCT-related publications, some arbitrary
choices had to be made.
 For instance, we shall use the term \emph{flow point} where some
references use \emph{flow-chart point}, \emph{program
  location} or \emph{function}, the latter obviously in a
functional-programming context.

Throughout the text, the symbol $\rhd$ will serve as a meta-variable ranging
over relations $\{>, \ge\}$. 

\subsection{Monotonicity constraint systems: structure} 
\label{sec:mcs}

A monotonicity constraint system is an abstract program.
An abstract program is, essentially, a set of \emph{abstract transitions}. An
abstract transition is a relation on (abstract) program states. 

When describing program transitions, it is customary to mark the variables
in the \emph{resulting} state with primes (e.g., $x'$). 
For simplicity, we will name the variables $x_1,\dots,x_n$ (regardless of
what flow point we are referring to).
Of course, 
in practice there is no reason for the number of variables to be the
same throughout the program, but this does not affect the theory in any
essential way.

\bdfn 
A \emph{monotonicity constraint system}, or MCS, is an abstract
program representation that consists of a control-flow graph (CFG),
monotonicity constraints and state invariants, all defined below.
\begin{enumerate}[$\bullet$]
\item
A control-flow graph is a directed graph (allowing parallel arcs)
over the set $F$ of \emph{flow points}.

\item
A \emph{monotonicity constraint} (MC) 
is a conjunction of 
order constraints $x \rhd y$ where
 $x,y\in\{x_1,\dots,x_n,x_1',\dots,x_n'\}$. 

\item
Every CFG arc $f\to g$  is associated with a
monotonicity constraint $G$. We write $G:f\to g$.

\item
For each $f\in F$, there is an \emph{invariant} $I_f$, which is
a conjunction of order constraints among the variables $\{x_1,\dots,x_n\}$.
\end{enumerate}
\edfn

In writing order constraints, we will also use $<,\le$ as a syntactic sugar
($x<y$ is $y>x$), and $x=y$ to mean $x\ge y \land y\ge x$.

The terms ``abstract program", ``constraint system" and ``MCS instance" are used
interchangeably, when context permits. The letters $\cal A$, $\cal B$ are usually used
to denote such a program; $F^{\cal A}$, $F^{\cal B}$ denote their respective flow-point sets. 

We next show how an MC is represented by a labeled digraph (directed graph); 
the notation $x\xrightarrow{r} y$ represents an arc  from $x$ to $y$ with label $r$.

\bdfn[constraints as graphs]
The graph representation of a monotonicity constraint is a labeled digraph $(V, E)$
with $V = \{x_1,\dots,x_n,x'_1,\dots,x'_n\}$ and
$E$ includes a labeled arc for each constraint: specifically,
  for a constraint $x\rhd y$, an arc $x\xrightarrow{\rhd} y$.

\ifLMCS\else\noindent\fi
The labeled arcs are referred to, verbally, as strict (for $>$) and non-strict ($\ge$).
\edfn

Note that arcs may connect two source variables, two target variables
or a source and a target variable---in any direction.
The notation $x\to y$ may be used to represent an arc from $x$
to $y$ (of unspecified label). In diagrams, to avoid clutter, we distinguish
the types of arcs by using a dashed arrow for the weak inequalities or 
equalities (see Figure~\ref{fig:excls}).

An equality constraint $x=y$ is represented by a pair of non-strict arcs. In certain algorithms,
it is convenient to assume that the such arcs are distinguished from ``ordinary''
non-strict arcs. We refer to them as \emph{no-change arcs}.

Henceforth, we identify a MC with its graph representation:
 it is a graph and a logical conjunction of constraints at the same time.
Context will usually clarify what view is taken,.

\bex \label{ex:cls} 
Consider the following (contrived) program fragment:
\begin{verbatim}
while (x,y,z > 0)
   if (y>x) 
      y = z;   x = unknown();  z = x-1
   else 
      z = z-1; x = unknown();  y=x-1
\end{verbatim}

The program computes over non-negative integers, which justifies the well-founded model.
For representing this program as an MCS in an economic way, we transform each
basic block (namely each branch of the {\tt if}) into an MC.  The CFG thus consists of
a single flow-point, representing the top of the loop, with two self-loops, $G_1$ and
$G_2$:
\begin{align*}
G_1&:\quad
 \pgt{x}<\pgt{y}\land \pgt{z}=\pgt{y}' \land \pgt{x}'>\pgt{z}' 
\\
G_2&:\quad
 \pgt{x}\ge\pgt{y}\land \pgt{z}>\pgt{z}' \land \pgt{x}'>\pgt{y}' 
\end{align*}
Figure~\ref{fig:excls} shows the graph representation of this MCS.
\eex
\edef\exCLSstart{\thepage}

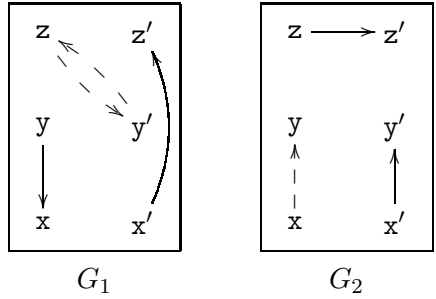
\begin{fig0}{MCS example.
The CFG (not shown) consists of a single flow-point and two self-loops.
In the MC graphs, the left-hand side is the source. Broken arcs represent non-strict descent.}{t}
{fig:excls}
\begin{centering}
\setlength{\extrarowheight}{1ex}
\begin{tabular}{@{\extracolsep{20pt}}cc}
\fbox{\ $\xymatrix@R=20pt{
  \texttt{z}\ar@/_4pt/@{-->}[dr] & \texttt{z}' \\
  \texttt{y}\ar@{->}[d]  & \texttt{y}'\ar@/_4pt/@{-->}[ul] \\
  \texttt{x}        & \texttt{x}'\ar@/_10pt/@{->}[uu]
}$\ } &
\fbox{\ $\xymatrix@R=20pt{
  \texttt{z}\ar@{->}[r] & \texttt{z}' \\
  \texttt{y}  & \texttt{y}' \\
  \texttt{x}\ar@{-->}[u]        & \texttt{x}'\ar@{->}[u]
}$\ } \\
$G_1$ & $G_2$  \\
\end{tabular}\par
\end{centering}
\end{fig0}

One may wonder why only conjunctions are allowed. In fact, 
any Boolean combination of constraints can be put into disjunctive
normal form, and then split into several MCs, one for each disjunct.
Of course, there are combinations (e.g., in conjunctive normal form) that
will be blown up exponentially. Apparently, this has not a problem in 
applications of SCT so far. 

Another natural question is why state invariants are included in the
definition, as it is
possible to include the constraints $I_f$ in every MC that transitions from $f$,
making the state invariant redundant. However, it may be convenient (and
is very much so in the algorithms of 
Sections~\ref{sec:elaborate}--\ref{sec:grf}) to make the
association of certain constraints with a flow point, rather than a 
specific tansition, explicit. Note also that static analysis
algorithms (e.g., inverval analysis~\cite{CousotHalbwachs:1978})
 often associate invariants with locations in the program, taking
into account all transitions that lead to a given location.

The size-change graphs of~\cite{leejonesbenamram01} are a simple class of
monotonicity constraints:

\bdfn[size-change graph] \label{def-scg}
A size-change graph (SCG) is a monotonicity constraint
consisting only of
relations of the forms $x_i\rhd x_j'$. 
 As a graph, it is bipartite and includes only arcs from
source variables to target variables.

An \emph{SCT instance} is a MCS where all constraints are size-change graphs
and all invariants are trivial.
\edfn

\subsection{Semantics and termination}
\label{sec:semantics}

Recall that a well-order is a total order with no infinite strictly-descending
chains, in other words, a well-founded total order.
The semantics of the
abstract program assumes a well-ordered set
$\val$ as the domain for variables' values. The non-strict order relation on
$\val$ is $\ge$ and its strict part is denoted by $>$ (that is,
$x>y \iff x\ge y \land x\neq y$).

While it seems that most applications of SCT use total orders (e.g., comparing
data objects by their size), there are exceptions (notably in conjunction
with term rewriting systems). In fact,
all notions and results in this paper
work equally well with partial orders, and even partial quasi-orders (where $\ge$
is not antisymmetric), except for the global ranking functions,
which presume a total order.  For uniformity, I chose to make total order the
basic assumption.

\bdfn[states]
Let $\cal A$ be an $n$-variable MCS.
A {\em state\/} of $\cal A$ (or an abstract state)
is a pair $(f,\sigma)$, where
$f\in F^{\cal A}$ and $\sigma:\{1,\dots,n\}\to\val$ represents
 an assignment of values to the variables.
\edfn

Satisfaction of a predicate $e$ with free variables $x_1,\dots,x_n$ (for example, $x_1>x_2$) by
an assignment $\sigma$ is defined in the natural way, and expressed by
$\sigma\models e$.
If $e$ is a predicate involving the $2n$ variables $x_1,\dots,x_n, 
x_1',\dots,x_n'$, we write $\sigma,\sigma' \models e$ when $e$ is
satisfied by setting the unprimed variables according to
$\sigma$ and the primed ones according to $\sigma'$.

\bdfn[transitions] \label{def:trans}
 A transition is a pair of states, a \emph{source state} $s$ and a \emph{target state} $s'$.
For $G:f\to g\in {\cal A}$, we write $\trans{G}{(f,\sigma)}{(g,\sigma')}$ if
$\sigma\models I_f$, $\sigma'\models I_g$ and $\sigma,\sigma' \models G$.
We say that transition $(f,\sigma)\mapsto
 (g,\sigma')$ is \emph{described by $G$}.
 
$G$ is called \emph{unsatisfiable} if it describes no transition.

The \emph{transition system} associated with ${\cal A}$
is the relation $T_{\cal A}$
 defined by $$(s,s')\in T_{\cal A} \,\iff\, \trans{G}{s}{s'}
\mbox{ for some $G\in {\cal A}$}.$$
\edfn

Practically, the transition system would be an abstraction of the transitions
 of a concrete
program. The variables of flow points might
represent actual data in the program, but
quite often they are already an abstraction, like the size of a concrete object
(this is, in fact, the source for the name of the SCT method).

\bdfn[run]
A {\em run\/} of ${\cal T}_{\cal A}$ is a (finite or infinite) sequence
of states $\tilde s = s_0,s_1,s_2\dots$ such that for all $i$,
 $(s_i, s_{i+1})\in {\cal T}_{\cal A}$.
\edfn

Note that by the definition of ${\cal T}_{\cal A}$, a run is associated
with a sequence of CFG arcs labeled by $G_1,G_2,\dots$ where
$\trans{G_i}{s_{i-1}}{s_i}$. This sequence constitutes a walk in the CFG
(recall that a walk is a directed path where repeated nodes and arcs are
allowed).

\bdfn[termination]
Transition system ${\cal T}_{\cal A}$ is {\em uniformly terminating\/} if
it has no infinite run.
\edfn

MCS $\cal A$ is said to be \emph{terminating} if
${\cal T}_{\cal A}$ is uniformly terminating for any choice of $\val$.


\bdfn[$\vdash$]  \label{def:vdash}
Let $P(s,s')$ be any predicate over
states $s,s'$, possibly written
using variable names, e.g., $x_1>x_2 \land x_2<x_2'$.
We write $G\vdash P$ if
$\forall s,s': \trans{G}{s}{s'} \Rightarrow P(s,s')$.
\edfn

\bdfn
A \emph{global ranking function} for a transition system $\cal T$
with state space $\mathit St$
is a function $\rho:{\mathit St}\to W$, where $W$ is
a well-founded set, such that $\rho(s) > \rho(s')$ for every $(s,s')\in {\cal T}$.

A ranking function for a MCS $\cal A$ is a 
ranking function for $T_{\cal A}$. Namely, it
satisfies $G\vdash \rho(s) > \rho(s')$ for every $G\in {\cal A}$.
\edfn

The qualifier \emph{global} may be omitted in the sequel, since this paper does
not deal with the notion of local ranking functions.

\subsubsection*{Example \ref{ex:cls}} (continued).
The MCS shown on Page~{\exCLSstart}
is terminating. To see this, note first that 
 the set of constraints
$\trans{G_2}{s}{s'}\land \trans{G_1}{s'}{s''}$ is unsatisfiable
(the reader is invited
to verify this by writing the constraints down, or by inspecting the graphs).
Thus, a $G_2$-transition
will never be followed by a $G_1$; in other words, a valid run can switch from $G_1$
to $G_2$ but cannot switch back.  In $G_2$, variable $\pgt{z}$ decreases;
it also decreases in $G_1$, \emph{if followed by $G_1$ again} (the reader is, again, invited
to verify this).

We conclude that the following
function descends (lexicographically) in every possible transition, and constitutes
a ranking function for this program:
\[ \rho(\pgt{x},\pgt{y},\pgt{z}) = \left\{\begin{array}{cl}
        \langle 1,\pgt{z}\rangle & \mbox{if $\pgt{y}> \pgt{x}$} \\
        \langle 0,\pgt{z}\rangle & \mbox{if $\pgt{y} \le \pgt{x}$}. 
\end{array}\right.
\]

Note that the MCs are not size-change graphs. Their best approximations
as size-change graphs are $G_1^{\text{scg}} = \{\pgt{z} > \pgt{y}'\}$ 
and  $G_2^{\text{scg}} = \{\pgt{z} > \pgt{z}'\}$,
which do not prove termination.
In fact, the issue of unsatisfiable combinations of constraints
(as for $G_2$ followed by $G_1$)
never arises with SCT instances;
this is one of the salient differences between the MC abstraction and the SCT one.

\section{Multipaths, Walks and Termination}

The purpose of this section is to formulate the \emph{combinatorial} termination condition
for monotonicity constraint systems, that is, the property that MC graphs should satisfy
so that termination of any associated transition system can be deduced.  This approach
was very useful in the study of the SCT abstraction, and we would like to define the
corresponding notions for MCS so that for the special case of SCT, they match the known
results.

The size-change termination principle~\cite{leejonesbenamram01} states that a program is known
to terminate if ``in every (hypothetic) infinite run, something descends infinitely''.
 In order to analyse such an infinite run, we analyse sequences of size-change
graphs (and more generally, MC graphs) that describe such runs. 
For this purpose, we introduce the
concept of a \emph{multipath} (following~\cite{leejonesbenamram01}).

\begin{fig0}{A multipath.}{t}{fig-multipath}
$$\xymatrix@R=20pt{
  x[0,1]\ar@/_4pt/@{-->}[dr] & x[1,1] \ar@{->}[r] & x[2,1]\ar@/_4pt/@{-->}[dr] & x[3,1]\\
  x[0,2]\ar@{->}[d]  & x[1,2]\ar@/_4pt/@{-->}[ul]  & x[2,2]\ar@<1ex>@{->}[d]  & x[3,2]\ar@/_4pt/@{-->}[ul]\\
  x[0,3]        & x[1,3]\ar@<-1ex>@/_10pt/@{->}[uu]\ar@{-->}[u]   & x[2,3]\ar@<4pt>@{->}[u]& x[3,3]\ar@<-1ex>@/_10pt/@{->}[uu]
}$$
\end{fig0}

\subsection{The Criterion for Monotonicity Constraint Systems}
\label{sec:criterion}

\bdfn[multipath]
Let $\cal A$ be an $n$-variable MCS, and let 
$f_0\stackrel{G_1}{\to}f_1\stackrel{G_2}{\to}f_2\ldots$ be an MC-labeled path
in the CFG.
The {\em multipath} $M$ that corresponds to this path
is a (finite or infinite) graph with nodes $x[t,i]$, where $t$ ranges from 0
up to the length of the path,
and $0<i\le n$. Its arcs are obtained by merging the following sets:
for all $t\ge 1$, $M$ includes the arcs of
$G_t$, with source variable $x_i$ renamed to $x[t-1,i]$ and target variable
$x_j'$ renamed to $x[t,j]$. In addition, for all $t\ge 0$, arcs representing
the invariant $I_{f_t}$ are included (with node $x[t,i]$ representing $x_i$).

The multipath may be written concisely as $G_1G_2\dots$; if $M_1, M_2$ are
finite multipaths, $M_1$ corresponding to a CFG path that ends where $M_2$
begins, we denote by $M_1M_2$ the result of concatenating them
in the obvious way.
\edfn

Figure~\ref{fig-multipath} depicts multipath $G_1G_2 G_1$, based on the MCs from Figure~\ref{fig:excls}.

Clearly, a multipath represents a conjunction of constraints
on a set of variables associated with its nodes. 
We can consider assignments $\sigma$ to these variables, where the value
assigned to $x[t,i]$ may be
denoted $\sigma[t,i]$; such an assignment may satisfy the constraints,
or not.
A satisfying assignment defines a concrete run of ${\cal T}_{\cal A}$,
along the given CFG path.

If we start at any node in a multipath, and walk along arcs, we are tracing a
descending chain of values. This way, walks in the multipath may be used to prove
when infinite multipaths are unsatisfiable.

\bdfn
A walk that includes a strict arc  is said to be \emph{descending}.
A walk that includes infinitely many strict arcs  is \emph{infinitely descending}.
\edfn

\bdfn[termination criterion] \label{def:MCSterm}
An MCS  $\cal A$ is \emph{size-change terminating} if every 
infinite $\cal A$-multipath has an infinitely descending walk.
\edfn

Note that the walk above may actually be a cycle! In this case it is contained
in a finite section of the multipath, and, logically, it implies the condition
$x>x$ for some variable $x$. Thus, such a multipath is unsatisfiable and
corresponds to no concrete run. If the walk is not a cycle, it indicates
an infinite descending chain of values and this contradicts well-foundedness.
Once again, we conclude that the multipath is unsatisfiable.
We conclude that if $\cal A$ is size-change terminating, it can have
no infinite runs. 

\blem \label{lem:sound}
If MCS $\cal A$ is size-change terminating, it is (semantically) terminating.
\elem

\bprf
Suppose that $\cal A$ is size-change terminating.
For any (hypothetical) infinite run $\tilde s$
of $T_{\cal A}$ there is an underlying infinite path in the CFG, which
induces a
multipath $M$; the values assigned to variables in $\tilde s$ should satisfy
all constraints expressed by $M$, which is impossible because of the infinitely
descending walk.

\eprf

Thus, size-change termination is a sound criterion for termination.

To show completeness, we suppose that an infinite multipath without infinite
descent exist, and show the existence of a domain $\val$ 
over which this infinite multipath is satisfiable.
This is achieved using the following definition and lemma.

\bdfn
Let $S$ be any set. A binary
relation $\succ$ on $S$ is a \emph{strict order}
if it is transitive and irreflexive. A binary relation $\succeq$ is 
a \emph{non-strict quasi-order} if it is transitive and reflexive;
if it is also antisymmetric, it becomes a (partial) order---the word 
\emph{partial} tacitly applies to all the above. 

Let $\succ$ be a strict order and $\succeq$ a non-strict
quasi-order.
The relations are \emph{compatible} if
\begin{eqnarray}
a \succ b &\Rightarrow&   a \succeq b \\     
a \succ b \land b\succeq c &\Rightarrow&   a \succ c \\
a \succeq b \land b \succ c &\Rightarrow&   a \succ c
\end{eqnarray}
\edfn




It seems to be ``folklore'' that a well-founded partial order can be extended to
a well-founded total order (given the axiom of choice or some similar machinery).
The next lemma makes a slightly stronger statement (involving a quasi-order and a compatible strict
order), and is given with proof\footnote{
The proof is inspired by a proof of the well-ordering theorem in
\url{http://planetmath.org/encyclopedia/ProofOfZermelosWellOrderingTheorem.html}
(dated 2007-03-12).}.

\blem \label{lem:wf}
Let $S$ be any set, $\succeq$ a quasi partial order on $S$, and $\succ$ a
compatible well-founded strict order.
There is a well-ordered set $B$ (with order $\ge$ and strict $>$) 
and a mapping $h:S\to B$ that agrees with the order relations, that is:
\begin{align*}
x \succeq y &\Rightarrow h(x)\ge h(y) \\
x \succ y &\Rightarrow h(x) > h(y) 
\end{align*}
\elem

\bprf
Let $X = {\cal P}(S) - \emptyset$. For each $x\in X$,
let $M_x$ be the set of minimal elements of $x$ under $\succ$. This is always non-empty.
For $x\in X$ and $a\in x$, let $U^x_a = \{b\in x\mid b\preceq a\}$; note that 
if $a\in M_x$ then $U^x_a\subseteq M_x$
and $a\in U^x_a$, so $U^x_a$ is not empty.

Using the Axiom of Choice,
let $f$ be a choice function with $f(x)\in M_x$ for all $x$. Define, by transfinite induction over 
\emph{the class of all ordinals}, the partial function with range ${\cal P}(S) \setminus \{\emptyset\}$:
\begin{eqnarray*}
g(\beta) = U^y_{f(y)}, &\mbox{ where }&
y = S - \bigcup_{\gamma<\beta} g(\gamma),\\ &\mbox{ unless }&
y=\emptyset \mbox{ or } g(\gamma) \mbox{ is undefined for some } \gamma<\beta.
\end{eqnarray*}

Define, by the ZF \emph{axiom of replacement}, 
$B =  \{ \beta \mid \exists x\in S : g(\beta) = x \}$.
Since $B$ is a set of ordinals, it cannot contain all ordinals (by the Burali-Forti paradox),
thus there is an ordinal $\alpha$ not in $B$. Identifying an ordinal with the set of smaller
ordinals, $(\alpha+1) \setminus B$ is a non-empty set of ordinals.
Since the ordinals are well ordered, there is a least ordinal $\beta\in (\alpha+1) \setminus B$;
in fact it is the least ordinal not in  $B$.
Therefore $g(\beta)$ is undefined. It cannot be that the second ``unless"
clause in the definition holds (since $\beta$ is the least such ordinal), so it must be
that $S - \bigcup_{\gamma<\beta} g(\gamma)  =\emptyset$,
 and therefore for every $a\in S$ there is some $\gamma<\beta$ such that $a\in g(\gamma)$.
In fact, by the definition of $B$, such $\gamma$ is unique, so
letting $h(a) = \gamma$ defines a total function $h:S\to B$.
 
We claim that $h$ agrees with the order relations.
To see this, let  $a,b\in S$.
 By the definition of $h$, 
$$a \in g(h(a)) \subseteq M_y \subseteq y = S - \bigcup_{\gamma<h(a)} g(\gamma).$$
Suppose that $a \succ b$.
If $h(a)\le h(b)$, then also $b\in y$, so $a$ is not minimal in $y$ and is not in $M_y$,
a contradiction. Thus, $h(a) > h(b)$, so $>$ agrees with $\succ$.
Next, suppose that $a\succeq b$; if $b\in y$, then 
(by the definitions of $M_y$ and $U^y_{f(y)}$), we have $b\in U^y_{f(y)} = g(h(a))$;
so $h(b)=h(a)$. If $b\notin y$, then $h(b)<h(a)$. Either way, $\ge$ is seen to agree with $\succeq$.
\eprf

\blem\label{lem:complete}
MCS $\cal A$ is terminating only if it is size-change terminating. 
\elem
\bprf
 Suppose that $\cal A$ is not size-change terminating.
Hence, an infinite
multipath $M$ can be formed, without infinite descent. 
Let ${S}$ be the set of nodes of $M$, identified by
the usual notation $x[t,i]$. Our aim is to show that there is a model for $M$,
that is, $M$ is satisfiable.

 We first define order relations on $S$.
Specifically, define the relation $\succeq$ on $S$ by: 
 $x[t,i]\succeq x[s,j]$ if and only if
 $M$ includes a walk from $x[t,i]$ to $x[s,j]$.
Define the relation $\succ$ on $S$ by: 
 $x[t,i] > x[s,j]$ if and only if
 $M$ includes a descending walk from $x[t,i]$ to $x[s,j]$.
 
It is easy to verify that $({S},\succeq,\succ)$ satisfy the assumptions of the
last lemma.  We choose $\val$ to be the well-ordered set $B$ from the conclusion of the lemma,
and $\sigma$ to be the mapping $h$; then $\sigma$ is an assignment that satisfies $M$.
\eprf

\noindent
Combining Lemmas~\ref{lem:sound} and~\ref{lem:complete}, we have
\bthm \label{thm:soundandcomplete}
MCS $\cal A$ is terminating if and only if it is size-change terminating. 
\ethm

\subsection{The SCT criterion}

The SCT condition~\cite{leejonesbenamram01} is similar to the MCs termination
condition (Definition~\ref{def:MCSterm}), but only concerns walks that proceed forward in
the multipath. Obviously, with SCT graphs, there are no other walks anyway.

\bdfn
In a multipath, a thread is a walk that only includes arcs 
in a forward direction
($x_i\to x_j'$).
%
We say that MCS $\cal A$ \emph{satisfies SCT} if every 
infinite $\cal A$-multipath has an infinitely descending thread.
\edfn

Note that for a general MCS, there is a difference between being size-change terminating
and ``satisfying SCT." The latter refers to the criterion defined above.

Example~\ref{ex:cls}  shows that the SCT condition, 
while clearly a sufficient condition for termination, is not a necessary one
when general monotonicity constraints are considered.

\section{Deciding Termination: the Closure Algorithm}
\label{sec:closure}

We will show in this section an algorithm to decide MCS termination. 
This algorithm is very similar to algorithms used in previous works such as
\cite{Sa:91,LS:97,DLSS:2001} which, however, are not complete
(see Section~\ref{sec:compare}).

\subsection{Consequence-closure and composition}

\bdfn
A monotonicity constraint $G$ is \emph{closed under logical consequence}
(or just closed) if, whenever
$G\vdash x > y,$ for 
$x,y\in \{x_1,\dots,x_n, x_1',\dots,x_n'\}$,
this constraint is explicitly included in $G$;
and if $G\vdash x \ge y,$ but not $x>y$, then $x\ge y$ is included in $G$.
\edfn

Note that for $G:f\to g$,
the condition $G\vdash P$ takes the invariants $I_f$
and $I_g$ into account (consider Definitions~\ref{def:trans} and~\ref{def:vdash}).
Thus, a closed MC subsumes the invariants in its source and target
states.

\bdfn
An MC $H$ \emph{is at least as strong as} $G$
if whenever $G\vdash P$, also $H\vdash P$.
The (consequence) \emph{closure} of a monotonicity constraint $G$,
denoted $\overline G$, is 
the weakest MC
that is at least as strong as $G$ and is consequence-closed.
\edfn

Practically,
calculating the closure means inserting into $G$ all the relations that
can be deduced from $G$. This 
is easy given the graph representation of the MC, as described next.

\bdfn \label{def:floyd}
The \emph{weighted-graph representation} of an MC $G:f\to g$ consists of the same
nodes as $G$, and an arc $x\to y$ for every relation $x\rhd y$ included
in $G$, $I_f$ or $I_g$\footnote{Parallel arcs can be eliminated, 
preferring $>$ to $\ge$.}.
Each non-strict arc is given a weight of 0 and each strict arc, a weight of $-1$.
\edfn

\blem \label{lem:floyd}
Let $G^{wgt}$ be the weighted-graph representation of MC $G$.
Then $G$ is unsatisfiable if and only if a negative-weight cycle exists
in $G^{wgt}$.
Assuming that $G$ is satisfiable, 
$\overline G$ includes an arc $x\to y$ if and only if a path
from $x$ to $y$ exists in $G^{wgt}$; this arc is strict if there is a path of
negative weight.
\elem

We omit the straight-forward justification of this lemma.
It implies that a standard 
All-Pairs Lightest-Path (weighted shortest-path) algorithm, such as
the Floyd-Warshall algorithm~\cite{CLRS},
can be used to find if $G$
is satisfiable, and compute $\overline G$ if it is, in polynomial time---%
specifically, $O(n^3)$. 

Remark: this is not best possible, asymptotically, but is
a simple solution. Different improvements may be tried, but whether they
are of practical value cannot be judged on a theoretical level, and would only make
sense if the graphs are large. For example, one may divide the running time 
by a constant, dependent on the machine word length, using 
a bit-level representation. For
sparse graphs with $a \ll n$ arcs, one may opt for an $O(na)$ solution 
based on repeated DFS search (details are left to the reader).

\bdfn[composition] \label{def-composition}
The {\em composition} of MC $G_1:
f\to g$ with $G_2: g\to h$, written $G_1;G_2$, is a MC with source
{$f$} and target {$h$}, which includes 
all the constraints among $s,s'$ implied by
$\exists s'' : \trans{G_1}{s}{s''}\land \trans{G_2}{s''}{s'}$.
\edfn

Composition, too, can be implemented efficiently by a lightest-path
algorithm applied to
the multipath $G_1G_2$.
As already noted, the procedure also determines whether the multipath
is satisfiable (note that, as Example~\ref{ex:cls} demonstrates,
two satisfiable MCs may form an unsatisfiable multipath).

The same graph procedure can be applied to any finite multipath 
$M = G_1\dots G_\ell$.
 It computes $\overline M = \overline{G_1;\cdots;G_\ell}$, which we call the
\emph{collapse} of $M$ (if $\ell =1$, it is the consequence-closure of $G_1$).
Thus, $\overline M$ includes an arc $x_i\to x_j'$ if and only if $M$ includes a
path from node $x[0,i]$ to node $x[\ell,j]$; and similarly for the other combinations
($x_i\to x_j$, $x_i'\to x_j'$ and $x_i' \to x_j$).
The arc is strict if and only if the path includes a strict arc. 
 This gives us the following observation

\begin{obs} \label{obs:segments}
Consider an infinite $\cal A$-multipath $M$, represented as
the concatenation of finite segments $M_1M_2\dots$, 
and let ${M}' =
(\overline M_1)(\overline M_2)\dots$
\mbox{If ${M}'$} has an infinite descending walk, so does $M$.  
\end{obs}


\bdfn[closure set]
Given an MCS $\cal A$,  its closure set ${\cal A}^*$ is
$$\{\overline M \mid M \mbox{ is a satisfiable $\cal A$-multipath}\}. $$
\edfn

\noindent
The set ${\cal A}^*$ is finite, since there are finitely many possible MCs. 

\subsection{A termination test}

\bdfn[cyclic multipath]
We say that a multipath $M$ (possibly a single MC)
is \emph{cyclic} if its source and target flow-points are equal.
This is equivalent to stating that $MM$ is a valid multipath.
\edfn

\bdfn[circular variant]
For an MC $G$, 
the {\em circular variant\/} $G^\cv$ of $G$ is a directed graph
obtained by adding, for every parameter $x_i$, an edge
$x_i  \edge x_i'$. This edge is treated as a pair of no-change arcs,
but is distinguished from any arcs already present in $G$. These
additional edges are called \emph{shortcut edges}.
\edfn

This definition is meant to be used with a cyclic MC.  The shortcut edges are used
to analyze the effect of juxtaposing multiple copies of $G$ (as in the multipath $GG$).
This will be clearer in the proof of Theorem~\ref{thm-graph-condition}.

\bdfn[types of cycles]
Let $G$ be a cyclic MC.
A  cycle in $G^\cv$ is a  path commencing and ending at
the same node.  It is a \emph{forward cycle}
if it traverses shortcut edges more often in the backward
direction (from $x_i'$ to $x_i$) than it does in the forward direction%
\footnote{this naming may seem strange, but will will later
see that such a cycle is ``unwound'' into a forward-going walk in the
multipath $G^\omega$.}.

A \emph{balanced cycle} is a cycle
that traverses shortcut edges equally often in both
directions.

A cycle is \emph{descending} if it includes a strict arc.
\edfn

\bdfn[Local Termination Test]
We say that $G$ passes the \emph{Local Termination Test}, or LTT, if $G$ has a
descending cycle, either forward or a balanced.
\edfn

See Figure~\ref{fig:ltt}, Parts (a)-(c), for an example of a forward
descending cycle. 	If arc $x_3' \nonstrict x_1'$ were
replaced with $x_3' \nonstrict x_2'$, a balanced cycle would ensue.

Using the graph representation, 
the Local Termination Test in not hard to implement;
let us sketch the algorithm, leaving the proof and fine details
to the interested reader. 
First, we note that one can test one strongly connected component (SCC)
of $G^\cv$ at a time, and only if it includes a strict arc. Secondly,
we claim that 
if there is any forward cycle in the SCC, then it contains
a forward descending cycle.
This case can be identified by assigning weights to arcs and looking for a negative-weight
cycle using, say, the Bellman-Ford shortest-path
algorithm~\cite{CLRS}; this takes
$O(na)$ time if the graph has $n$ nodes and $a$ arcs\footnote{%
A better result, at least theoretically, is $O(an^{1/2})$ for $n$
nodes and $a$ arcs, due to Goldberg~\cite{Goldberg1995}.}.
  If no cycle of this kind exists,
there can only be a balanced descending cycle,
which would constitute a zero-weight cycle; 
having computed the single-source shortest-path distances,
such cycles can easily be found since they are cycles in the
shortest-path graph.

\bthm\label{thm-graph-condition}
MCS $\cal A$ is size-change terminating if and only if
 every cyclic MC in ${\cal A}^*$ passes the Local Termination Test.
\ethm

\bprf
For the forward implication (the ``if''), suppose that
 every cyclic MC in ${\cal A}^*$ passes the Local Termination Test.
Consider an infinite $\cal A$-multipath
${M}=G_1G_2\ldots$ and assume, by way of contradiction, that it is
satisfiable.

Consider the set of positive integers,
and label each pair $(t,t')$ , where $t<t'$, by 
\[
G = \overline{G_{t}; G_{{t+1}}; \cdots G_{{t'-1}}}
\]
which must be in ${\cal A}^*$, since this multipath is satisfiable.
Thus every pair has a label, and the set of labels is finite.
By Ramsey's theorem (in its infinite version), there is an infinite set of
positive integers, $I$, such that all pairs $(t,t')$ with $t,t'\in I$
carry the same label $G_I$.

Thus for any $t,t'\in I$ with $t<t'$,
$\overline{G_{t}; G_{{t+1}}; \cdots G_{{t'-1}}} = G_I$.
By Observation~\ref{obs:segments}, it now suffices to show that multipath
$(G_I)^\omega$ (infinite sequence of $G_I$'s) has an infinite descending walk.

Let $v_1,e_1,v_2,e_2,\dots,e_{s-1},v_s$ be the nodes and arcs (alternatingly)
 of the descending cycle
in $G_I$, where each $v_j$ is either $x_{i_j}$ or $x_{i_j}'$ for some index
$i_j$.  We can map the
cycle onto a walk in $(G_I)^\omega$, as follows. The first node is
$x[s,i_1]$. If the arc $e_1$ is an ordinary arc of $G$, the walk follows this
arc to $x[s+1,i_2]$, $x[s,i_2]$ or $x[s-1,i_2]$ (depending on the direction of
the arc). If $e_1$ is a shortcut arc, the walk is not extended: $v_2$ is also
mapped to $x[s,i_1]$ (necessarily $i_2=i_1$). We proceed in this 
manner until we complete the cycle and return to $v_1$. At this point,
our walk will have reached a node $x[s',i_1]$ for some $s'$. The relation among the forward
and backward crossings of shortcut edges implies that $s'\ge s$; specifically
$s'=s$ in the ``balanced'' case, and $s'>s$ in the ``forward'' case (hence the name).
 In the former case, we have discovered a descending cycle in
$(G_I)^\omega$ which indicates unsatisfiability. In the latter, we have a walk from
$x[s,i_1]$ to $x[s+d,i_1]$ for some positive $d$; another such walk can be
added to reach $x[s+2d,i_1]$, and so on; we conclude that an
 infinitely-descending walk occurs in $(G_I)^\omega$.
This establishes the forward implication (see Figure~\ref{fig:ltt}).

\begin{figure}[t]
\begin{centering}
\begin{tabular*}{0.9\textwidth}{@{\extracolsep{\fill}}ccc@{\extracolsep{0pt}}c}
$\xymatrix@R=10pt{
  x_1\ar@{->}[dr] & x_1' \\
  x_2\ar@{->}[d]  & x_2' \\
  x_3        & x_3'\ar@/_10pt/@{-->}[uu]
}$ &
$\xymatrix@R=10pt{
  x_1\ar@{->}[dr] & x_1'\ar@{<.>}[l] \\
  x_2\ar@{->}[d]  & x_2'\ar@{<.>}[l] \\
  x_3\ar@{<.>}[r] & x_3'\ar@/_10pt/@{-->}[uu]
}$ &
$\xymatrix@R=10pt{
 x_1\ar@{->}[dr] & x_1'\ar@{.>}[l] \\
  x_2\ar@{->}[d]  & x_2'\ar@{.>}[l] \\
  x_3\ar@{.>}[r] & x_3'\ar@/_10pt/@{-->}[uu]
}$ \\ 
 {\bf (a)} & {\bf (b)} & {\bf (c)} \\
\end{tabular*}\par
\end{centering}
\vspace{1cm}
\begin{centering}
\begin{tabular}{c}
$\xymatrix@C=15pt@R=20pt{
 x[0,1]\ar@{->}[dr] & x[1,1]\ar@{->}[dr] & x[2,1]\ar@{->}[dr] & x[3,1]\ar@{->}[dr]&\\
  x[0,2] & x[1,2]\ar@{->}[d] & x[2,2]\ar@{->}[d] & x[3,2]\ar@{->}[d] &\quad \dots\\
  x[0,3] & x[1,3]\ar@<-1ex>@/_10pt/@{-->}[uu]& x[2,3]\ar@<-1ex>@/_10pt/@{-->}[uu]& x[3,3]\ar@<-1ex>@/_10pt/@{-->}[uu] & \\
}$\\  {\bf (d)} \\
\end{tabular}\par
\end{centering}

\caption{(a) An MC $G$, (b) its circular variant $G^\circ$; (c) the directions of
shortcut edges are set  to form
a forward descending cycle, (d)
a walk in a prefix of $G^\omega$, corresponding to the cycle. 
The notation $x^t_i$ is a shorthand for $x[t,i]$.
(Example after
Codish, Lagoon and Stuckey)}
\label{fig:ltt}
\end{figure}

For the converse implication, suppose that MCS $\cal A$ is
size-change terminating, and let
$G$ be a cyclic MC in ${\cal A}^*$. Consider the multipath $G^\omega$.
Note that $G=\overline M$ for some finite, satisfiable $\cal A$-multipath $M$.
Let $\ell$ be the length of $M$.
Consider the infinite $\cal A$-multipath $M^\omega$, obtained by concatenating
infinitely many copies of $M$, to which we shall refer as \emph{blocks}.
By assumption, $M^\omega$ has an infinite descending walk.
If this walk is a cycle, contained entirely within one of the blocks,
 then $M$ is unsatisfiable, and $\overline M$ will not appear in ${\cal A}^*$.
 Thus, the walk must cross block boundaries.
Let us concentrate on the variables $x[t,i]$ occurring on these boundaries.
Since the walk is infinitely descending, some index $i$ has to occur twice
on the walk, say at nodes $x[t\ell,i]$ and $x[t'\ell,i]$ where $t\le t'$,
and there has to be 
a strict arc between these two points\footnote{%
These considerations apply equally to walks that are cycles and to ``open'' walks.}%
. There is thus a descending walk
from $x[t\ell,i]$ to $x[t'\ell,i]$.
Such a walk can be transformed into a cycle $C$ 
in $G^\cv$ in the following way:
starting with $x[t\ell,i]$, pick the first segment $S$ of the
walk (at least a single arc) that ends up again on a block boundary. If $ S$
ends up in $x[t+\ell,j]$ include in $C$
the $G$ arc $x_i\to x_j'$ followed by a backward
shortcut arc to $x_j$.
If $S$ ends up
in $x[t,j]$ for some $j\neq i$ then $G$ includes either an arc $x_i\to x_j$
or an arc $x_i'\to x_j'$. In the former case, include $x_i\to x_j$ in $C$;
in the latter,
include the path $x_i\to x_i'\to x_j'\to x_j$ where the first and last arcs
are shortcuts. Finally, if $S$
ends up in $x[t-\ell,j]$, include in $C$
the path $x_i\to x_i'\to x_j$ where the first arc
is a (backward) shortcut.
This process 
continues until we get to $x[t'\ell,i]$ in the original walk, whereupon $C$ will end at $x_i$,
becoming a cycle.
The inequality $t'\ge t$ implies that the number of backward shortcuts in $C$
will not exceed the number of forward shortcuts.

\ifLMCS\else\noindent\fi
We conclude that a descending cycle, either balanced or forward, exists in $G^\cv$.
\eprf

\begin{algo} (Deciding termination of an MCS ${\cal A}$) \label{alg-closure}
\end{algo}
\be[(1)]
\item Build ${\cal A}^*$ by a  transitive closure procedure:
\be[(a)]
\item Initialize a set $\cal S$ to $\{\overline G\mid G\in{\cal A}\mbox{ is satisfiable}\}$.

\item For any  $G:{ f}\to{ g}$  and  $H:{ g}\to{ h}$
 in     ${\cal S}$, include also $G; H$ in $\cal S$, unless it is unsatisfiable.
 
\item Repeat the above step is  until no more elements can be added to $\cal S$.
\ee
At this point, $\cal S$ is ${\cal A}^*$ (we omit a detailed proof).

\item
 For each  cyclic $G$ in $\cal S$, apply the Local Termination
Test to $G$. Pronounce failure (non-termination) if a graph that fails the
test is found.

\item 
 If the previous step has completed, the MCS terminates.
\ee

\subsection{Complexity of Algorithm~\ref{alg-closure}}

The determining factor in the worst-case complexity of the algorithm is the
number of elements in ${\cal A}^*$. An easy upper bound on the number
of possible MCs, ignoring the identity of the source and target flow-points,
 is $6^{2n^2}$,
as each MC has $2n$ variables and six possible relations among any 
pair of them ($x<y$, $x>y$, $x\ge y$, $x\le y$, $x=y$ or none).
Thus, the size of the closure set is at most $m^2 6^{2n^2}$, where $m$ is
the number of flow-points.

As long as the algorithm does not fail, any pair of elements that can be
composed (i.e., MCs associated with CFG arcs $f\to g$ and $g\to h$,
for any $f,g,h$) has to be considered
(Step 1.2).  To each such pair,
 we apply composition, and if the result is cyclic,
the local termination test. Thus we obtain an upper bound of
$$O(m^3(6^{2n^2})^2 n^3) = O(m^3 6^{4n^2} n^3)\,.$$
This is an over-estimate, however a significantly better bound (i.e., with $o(n^2)$ exponent)
is not known for this algorithm. Note that the next section provides algorithms
of better worst-case bounds: specifically, reducing the exponent from
$\Theta(n^2)$ to $\Theta(n\log n)$. 

We remark that the
management of data structures is, in this case, not costly, since all we need
is to maintain a set of MCs so that one can efficiently add an element while
testing to see if it was not already there. This can be done, for instance,
by a radix tree~\cite{CLRS}, where each operaton takes
linear time in the number
of bits that describe an MC (that is, $O(n^2)$).

Despite the exponential time and space complexity,
a similar algorithm for SCT has proved quite useful in practice;
and there are techniques to improve its performance, that are also applicable
here. Let us briefly discuss these (admitting that the ultimate test of
such techniques is empirical).

An important technique is 
reducing the size of the set $\cal S$ by
subsumption~\cite{BL:2006,FV:09}.
An MC $G$ is said to subsume $H$ if $G$ is less constrained than $H$. 
That is, every
transition described by $H$ is also described by $G$.
With consequence-closed graphs, testing subsumption is easy. And in this case,
one can safely ignore $H$. There is a cost involved in finding out whether
subsumption occurs, whenever an element is to be added to $\cal S$. But the
reduction in the size of $\cal S$ seems to make this worthwhile.

Another sort of optimization appears in~\cite{DLSS:2001,leejonesbenamram01}:
instead of applying the LTT to every element of ${\cal A}^*$, it suffices to do so for
the idempotent ones (elements $G$ such that $G;G=G$). This strategy does not reduce
the size of the closure set maintained, only the
 number of local tests. 
In SCT, it also reduces the complexity of the local test: as shown 
in~\cite{leejonesbenamram01}, it suffices to look for in-situ descent, that is,
an arc of the form $x_i\strict x_i'$ (which yields a forward
descending cycle of length 2).   This is not the case for general monotonicity
constraints; for a little example, note that the MC in Figure~\ref{fig:ltt} is idempotent, but does not include in-situ descent.

A warning is in order: if the set $\cal S$ is reduced by subsumption, it is
incorrect to test just the idempotent graphs. That is, there are non-terminating
instances where an idempotent counter-example will not be found, as subsumption
will have removed it%
\footnote{Alexander Krauss, private communication.}. 
Thus one has to choose between applying subsumption or testing only idempotent
graphs, and it seems clear that the former option has much greater impact.

\subsection{Complexity of the decision problem}
\label{sec:pspace}

What is the complexity class of the decision problem: \emph{MCS
Termination}? Algorithm~\ref{alg-closure} takes exponential time and space,
but a polynomial-space version is possible. It is most likely worst
in practice, but for theoretical completeness, let us prove

\bthm \label{thm:pspace}
The MCS Termination problem is PSPACE-complete.
\ethm

\bprf
It is known~\cite{leejonesbenamram01} that the SCT Termination problem
is PSPACE-hard, which also applies to MCS because SCT is a special case.
To show that the problem is in PSPACE, we will outline a non-deterministic
polynomial-space algorithm for the complement problem,
that is, non-termination.
The result will follow since (by Savitch's theorem)
$\mbox{coNPSPACE}=\mbox{NPSPACE}=\mbox{PSPACE}$.


Algorithm~\ref{alg-closure} can be seen
as a search for a counter-example---a cycle in the CFG that induces a
multipath that fails the test. The non-deterministic
algorithm guesses such a cycle. In each step, it
adds a transition to the cycle while computing the composition of the next MC
with a graph that represents the collapse of the
multipath traversed so far. Only this graph, along with the current flow-point,
have to be maintained in memory.  Whenever the current flow-point is the same
as the initial one, the local termination test is applied.
If at some point, an unsatisfiable MC results, the algorithm has failed to find
a counter-example. Otherwise it continues until finding one.
\eprf

\subsubsection*{Remark.}
Since both the MCS termination problem and SCT termination are PSPACE-complete,
a polynomial-time reduction of the former to the latter is known to exist.
However, it is not given explicitly, and it is not clear
whether such a reduction can be implemented efficiently enough to be actually useful.
On the other hand, the next section will show a reduction in
\emph{exponential time}, that, paradoxically, may be useful (in fact, the reduction produces
an \emph{easy instance} of SCT, one which can be analyzed in polynomial time,
as shown in Section~\ref{sec:grf}).

\subsection{Comparison to previous algorithms}
\label{sec:compare}

As previously mentioned, there have already been termination analyzers that
dealt with monotonicity constraints, but Codish et al.~\cite{Codish-et-al:05}
pointed out that their decision procedures were not complete; they emphasized
the case of integers (as was mentioned in the introduction), but the incompleteness
already occurs in the well-founded model.
In fact,
those procedures are similar to Algorithm~\ref{alg-closure}, and using the
terminology in this section we can present them so that the change in the
new algorithm becomes clear.

\bdfn
Let $G$ be a cyclic MC.
A  cycle in $G^\cv$ is a \emph{zig-zag cycle}
if it alternates arcs of $G$ with shortcut edges, the latter always
in the backward direction (from $x_i'$ to $x_i$).
\edfn

Note that a zig-zag cycle alternates forward arcs of $G$ ($x\to y'$)
with shortcut edges. This does not mean, however, that 
other arcs in the MC are ignored, because they might give rise to forward
arcs by transitivity.

\bdfn
We say that $G$ passes \emph{Sagiv's Test} if $G$ has a zig-zag
descending cycle.
\edfn

Clearly, this test is a special case of the LTT, and it should be easy to
see that they conincide for SCT graphs. The reader probably sees, already,
why this test is incomplete. In fact, Figure~\ref{fig:ltt} shows a terminating
instance that it misses.

Sagiv's test is from~\cite{Sa:91} and was used in the Termilog system
as described in~\cite{LS:97}, and the Terminweb system as described
in~\cite{CT:99}. Later, in~\cite{DLSS:2001}, the algorithm was
refined by observing that it suffices to test idempotent elements.
This improves its coverage (consider the cyclic MC:
$x>y \land x'<y'$) but it is still incomplete, by the same example.

It should be noted, that this comparison is based on a ``translation''
of the cited works to MCS terms. They actually analyze logic programs.
As for the algorithm proposed in~\cite{Codish-et-al:05}, the discussion  is
postponed to the next section where it fits more naturally.

\section{Fully Elaborated Systems and Stability}
\label{sec:elaborate}

In this section we describe a procedure that while increasing the size
of an abstract program, simplifies its termination
proof; in fact, we get back to the SCT condition. This result is
interesting theoretically, for understanding the relation between the two
formalisms, and also forms a basis for the algorithm in the following
section (that determines termination while constructing an explicit
ranking function).

The procedure duplicates flow-points while refining their invariants;
this means that computations that reach the same program location under
different conditions (i.e., different ordering of the variable values)
will be represented as reaching different flow-points
in the abstract program.
Similar transformations can be found in program
analysis in various guises (for example, in~\cite{LS:97} and subsequent works, flow points are queries to Prolog clauses
and are duplicated for different instantiation patterns).
 When we apply the transformation to monotonicity constraint
systems in a brute-force way,
we obtain what we shall call a \emph{fully elaborated system}.
We will see, in this section and the next, that fully elaborated systems are
MC systems of a particularly structured kind. 

In order to express the correctness of a transformation on abstract programs
we begin by defining ``simulation."

\subsection{Simulation}

We define the notions of simulation and bisimulation
for transition systems of the kind used in this work.
A transition system $\cal A$ simulates another one  $\cal B$
if (informally speaking) they have the same  transition sequences,
up to the identity of flow points and the indexing of variables.

\bdfn
Let $\cal A$, $\cal B$ be transition systems, with flow-point sets $F^{\cal A}$,
$F^{\cal B}$ respectively, 
and both having states described by $n$ variables over $\val$.
We say that $\cal A$ 
\emph{simulates}
$\cal B$  if there is a relation 
$\phi\subseteq F^{\cal B}\times F^{\cal A}$ (``correspondence of flow-points")
and, for all $(f,g)\in\phi$, a bijection
$\psi_{g,f}:  \{1,\dots,n\}\to\{1,\dots,n\}$  (``variable renaming")
 such that for every (finite or infinite) state-transition sequence
 $(f_1,\sigma_1)\mapsto (f_2,\sigma_2)\mapsto (f_3,\sigma_3)\mapsto \dots$ 
of $\cal B$ there is a corresponding sequence
$(g_1,\sigma'_1)\mapsto (g_2,\sigma'_2)\mapsto (g_3,\sigma'_3)\mapsto \dots$
of ${\cal A}$ with $(f_i,g_i)\in \phi$ and
$\sigma'_i = \sigma_i\circ(\psi_{g_i,f_i})$.
We say that $\cal A$ 
\emph{bisimulates} $\cal B$ if, in addition,
 for every (finite or infinite) state-transition sequence
$(g_1,\sigma'_1)\mapsto (g_2,\sigma'_2)\mapsto (g_3,\sigma'_3)\mapsto \dots$
of $\cal A$ there is a corresponding sequence
$(f_1,\sigma_1)\mapsto (f_2,\sigma_2)\mapsto (f_3,\sigma_3)\mapsto \dots$
of ${\cal B}$, also with $(f_i,g_i)\in \phi$ and
$\sigma'_i = \sigma_i\circ(\psi_{g_i,f_i})$.
\edfn

Thus, $\cal A$ bisimulates $\cal B$ if they simulate each other via the same
pair of mappings.

\bdfn
We say that an abstract program $\cal A$ (bi-)simulates an abstract program $\cal B$
if ${\cal T}_{\cal A}$ 
(bi-)simulates
${\cal T}_{\cal B}$, via mappings $\phi$ and $\psi$, as above.

We say that $\cal A$ simulates $\cal B$ \emph{deterministically} if for every
$f\in F^{\cal B}$ and assignment $\sigma$ satisfying $I_{f}$
there is a unique $g\in F^{\cal A}$ with
$(f,g)\in \phi$ such
that, letting $\sigma' = \sigma\circ(\psi_{g,f})$, assignment $\sigma'$ satisfies $I_{g}$.

If  $\cal A$ bisimulates $\cal B$, and  $\cal A$ simulates $\cal B$
deterministically, we say (for brevity) that $\cal A$ bisimulates
$\cal B$ deterministically.
\edfn

Determinism means that
the invariants of different $\cal A$
flow-points that simulate a given $\cal B$ flow-point have to
be mutually exclusive.

\begin{obs}
Suppose that $\cal S$ bisimulates $\cal T$.
Then $\cal S$ uniformly terminates if and only if $\cal T$ does.  
\end{obs}

\subsection{Elaboration}

\bdfn[full elaboration]
An MCS $\cal A$ is \emph{fully elaborated} if the following conditions hold:
\begin{enumerate}[(1)]
\item Each state invariant
fully specifies the relations among all variables. That is, for
$i,j\le n$, one of $x_i=x_j$, $x_i<x_j$ or $x_i>x_j$ is implied by $I_f$.

\item Each MC is closed under logical consequence.

\item No MC in $\cal A$ is unsatisfiable.
\end{enumerate}
\edfn

\newcommand{\lessoreq}{\left\{{<\atop =}\right\}}
Since the state invariant fully determines the relations among all
variables, we can \emph{re-index} the variables into sorted order,
so that the invariant becomes
\begin{equation}\label{eqn:order}
\textstyle x_{1} \lessoreq x_{2} \lessoreq\dots \lessoreq x_{n}.
\end{equation}
Of course, the re-indexing has to be incorporated also in
MCs incident to this flow-point, but this is straight-forward to do.
Indexing the variables in sorted order has some convenient consequences,
such as the having the property:

\bdfn \label{dfn:downwardclosure}
$G$ has the \emph{downward closure property}
if for all $k<j$, $x_{i}{\xrightarrow{>}} x_{j}'\in G$ entails
 $x_{i}{\xrightarrow{>}} x_{k}'\in G$; and $x_{i}{\xrightarrow{\ge}}
x_{j}'\in G$ entails
$x_{i}{\xrightarrow{\rhd}} x_{k}'\in G$ for some $\rhd\in\{>,\ge\}$.
\edfn


The number of possible orderings of $n$ variables plays a role in the combinatorics of
fully elaborated instances. Note that 
equalities are possible. Therefore, the number of orderings is not $n!$,
but a slightly larger number called \emph{the $n$th ordered Bell number} $B_n$.
An easily proved upper bound is $B_n\le 2n^{n-1}$
(consider two cases---no 
equalities, or at least one).
For more details, in~\cite[Seq.~A670]{Sloane}.
We denote the set of these orderings by $\bell_n$,
and assume that we fix some convenient representation so that 
``orderings'' can be algorithmically manipulated (for example, a
series of (in)equalities as in (\ref{eqn:order})).

The algorithm of full elaboration follows almost immediately
from the definitions.
\begin{algo} (full elaboration) \label{alg-fe}
Given an MCS ${\cal B}$, this algorithm produces a fully-elaborated MCS $\cal A$
that bisimulates $\cal B$.
\end{algo}
\be[(1)]
\item
For every $f\in F^{\cal B}$, generate flow-points 
$f_\pi$ where $\pi$ ranges over $\bell_n$.  
Define the variable renaming function $\psi_{f_\pi,f}$ so that
$\psi_{f_\pi,f}(i)$ is the $i$th variable in sorted order, according
to $\pi$. Thus,
$I_{f_\pi}$ will have exactly the form~(\ref{eqn:order}).
\item
Next, for every MC
$G:f\to g$ in $\cal B$, and every pair $f_\pi, g_\varpi$, create a size-change
graph $G_{\pi,\varpi}: f_\pi\to g_\varpi$ as follows:
\begin{enumerate}
\item
For every arc $x\to y\in G$, include the corresponding arc in $G_{\pi,\varpi}$,
according to the variable renaming used in the two $\cal A$ flow-points.
\item
Complete $G_{\pi,\varpi}$ by closure under consequences; 
unsatisfiable graphs (detected by the closure computation) are removed from the constructed system.
\end{enumerate}
\ee

\bex \label{ex:ELABORATION}
Let the system $\cal B$ consist of a single
 flow-point, say $f$, with $I_f=\texttt{true}$; and a single MC over two variables,
\[ G:\quad  x_1 > x_1' \land x_2 \ge x_2' \land x_1' \ge x_2'  \,.\]

In $\cal A$, we have $B_2 = 3$ flow-points $f_\pi$, namely:
$f_{[x_1<x_2]}, f_{[x_1=x_2]}, f_{[x_1>x_2]}$. For readability, let us denote the
variables in $\cal A$ by $y_i$ instead of $x_i$: then $y_i$ represents $x_i$ in
the first two flow-points, but in $f_{[x_1>x_2]}$, the indices are exchanged, to obtain
an increasing order of value.

Figure~\ref{fig:elaborate} shows three of the graphs $G_{\pi,\varpi}$, first when
initially constructed---just copying the arcs from $G$ according to the variable renaming,
then after adding the invariants 
 $I_{f_\pi}$ and $I_{f_\varpi}$, and finally after closure
 under consequences.
\eex

\begin{figure}[t]
\begin{centering}
\setlength{\extrarowheight}{1ex}
\begin{tabular}{l|c|c|c}
& $G_{[x_1<x_2],[x_1=x_2]}$ 
& $G_{[x_1<x_2],[x_1>x_2]}$ 
& $G_{[x_1=x_2],[x_1>x_2]}$ \\[1ex] \hline
Variables permuted &
$\xymatrix@R=20pt@C=20pt{
y_1\ar@{->}[r] & y_1' \ar@{-->}[d] \\
y_2\ar@{-->}[r] & y_2'\\
}$  &
$\xymatrix@R=20pt@C=20pt{
y_1\ar@{->}[dr] & y_1' \\
y_2\ar@{-->}[ru] & y_2'\ar@{-->}[u] \\
}$  &
$\xymatrix@R=20pt@C=20pt{
y_1\ar@{->}[dr] & y_1' \\
y_2\ar@{-->}[ru] & y_2'\ar@{-->}[u] \\
}$  \\\hline
 Invariants added &
$\xymatrix@R=20pt@C=20pt{
y_1\ar@{->}[r] & y_1' \ar@/_4pt/@{-->}[d] \\
y_2\ar@{-->}[r]\ar@{->}[u] & y_2'\ar@/_4pt/@{-->}[u] \\
}$  &
$\xymatrix@R=20pt@C=20pt{
y_1\ar@{->}[dr] & y_1' \\
y_2\ar@{-->}[ru]\ar@{->}[u] & y_2'\ar@{->}[u] \\
}$  &
$\xymatrix@R=20pt@C=20pt{
y_1\ar@{->}[dr]\ar@/_4pt/@{-->}[d] & y_1' \\
y_2\ar@{-->}[ru]\ar@/_4pt/@{-->}[u] & y_2'\ar@{->}[u] \\
}$  \\\hline
Closure &
$\xymatrix@R=20pt@C=20pt{
y_1\ar@{->}[r]\ar@{->}[dr] & y_1' \ar@/_4pt/@{-->}[d] \\
y_2\ar@{->}[r]\ar@{->}[u]\ar@{->}[ru] & y_2'\ar@/_4pt/@{-->}[u] \\
}$  &
$\xymatrix@R=20pt@C=20pt{
y_1\ar@{->}[dr]\ar@{->}[d]\ar@{->}[r] & y_1' \\
y_2\ar@{->}[ru]\ar@{->}[u]\ar@{->}[r] & y_2'\ar@{->}[u] \\
}$  &
$\xymatrix@R=20pt@C=20pt{
y_1\ar@{->}[dr]\ar@{->}[r]\ar@/_4pt/@{-->}[d] & y_1' \\
y_2\ar@{->}[ru]\ar@{->}[r]\ar@/_4pt/@{-->}[u] & y_2'\ar@{->}[u] \\
}$  \\
\end{tabular}
\end{centering}
\caption{Full-elaboration example. Dashed arcs are non-strict.}
\label{fig:elaborate}
\end{figure}

\subsubsection*{Complexity.}
 For an MCS $\cal B$, let $|{\cal B}|$ denote the number of abstract
 transitions (MCs) in ${\cal B}$ (without loss of generality, $|{\cal B}|
\ge |F^{\cal B}|$).
 
\blem \label{lem-refine}
Any MCS $\cal B$ with $n$ variables at any point
can be transformed into a fully-elaborated system $\cal A$,
deterministically bisimulating ${\cal B}$,
in $O(|{\cal B}| n^{2n+1})$ time and space. 
\elem

\bprf
This follows from a straight-forward analysis of Algorithm~\ref{alg-fe}.
Each MC $G$ yields $B_n^2 = O(n^{2n-2})$ offsprings $G_{\pi,\varpi}$, and the
work invested in each is $O(n^3)$ (as explained in the last section).
\eprf

\subsection{Stability}
\label{sec:stable}

The main result of this section is a proof that MCS termination can be reduced
to SCT termination, and that fully-elaborating the system achieves this.
We begin by formalizing the significant property, enjoyed by fully-elaborated systems,
which allows for the simplified termination proof: this 
property is called \emph{stability}.

\bdfn
An MCS $\cal A$ is \emph{stable} if (1) all MCs in $\cal A$ are closed under
 logical consequence;
(2) all MCs in $\cal A$ are satisfiable;
(3) For all $G:f\to g$ in $\cal A$, whenever
$G\vdash  x_i \rhd x_j$ (a relation between source variables),
that relation is included in $I_f$.
(4) Similarly, if
$G\vdash   x_i' \rhd x_j',$   
that relation is included in $I_g$.
\edfn

Note that just replacing every MC by its consequence-closure (if satisfiable) 
will satisfy conditions
(1) and (2), but will not necessarily make the system stable, since it may fail
to satisfy (3)-(4). 
In fact, these conditions may force flow-points to be duplicated, since two
MCs coming out of $f$ may disagree on the conditions that must be placed
in $I_f$.

\begin{obs}
A fully elaborated system is stable.
\end{obs}

 Full elaboration can be seen as a brute-force way of ``stabilizing''
 an MCS.
A system can also be stabilized by an iterative fixed-point computation,
which is likely to end up with less duplication of flow points and MCs.
For completeness, such an algorithm is described in 
Section~\ref{sec:stabilization}.
 But let us now first present the benefits of stability.

\blem \label{lem:stable=satisfiable}
In a stable MCS, every finite multipath is satisfiable.
\elem

\bprf
Let  $\cal A$  be a  stable MCS.
Let $M$ be a finite $\cal A$-multipath. For $M$ to be unsatisfiable, it must include
 a descending cycle.
We shall prove that if a descending cycle exists in $M$, then the
\emph{shortest} descending cycle must be contained in a single MC.
But this would make the MC unsatisfiable, contradicting the assumption of stability.

Suppose, to the contrary, that the shortest descending cycle spans more than one
MC.  Suppose that it spans MCs $G_{a},G_{a+1},\dots,G_{b}$.
We can asume that it includes a source node of $G_a$, for otherwise $G_a$ is
unnecessary (by stability, an arc among target nodes of $G_a$ also appears in $G_{a+1}$).
So, there is a
node $\nu = x[a-1,i]$ on the cycle. The node that precedes it on the cycle is a 
node of $G_a$, and so is the node that succeeds $\nu$.
Since all three nodes lie in the same MC, 
which is consequence-closed, this two-arc path can be replaced by a single arc,
deleting $\nu$ from the cycle.
Thus, the presumed shortest cycle is not shortest.
\eprf

\bthm \label{thm:stable=sct}
A stable MCS is terminating if and only if it satisfies SCT.
\ethm

\bprf
Let  $\cal A$  be a stable MCS.
If $\cal A$ satisfies SCT, it is terminating, since the SCT condition
is a special case of the MCS termination criterion.
For the converse direction, assume that $\cal A$ satisfies the MCS
termination criterion; we will prove that it satisfies SCT.
Let $M$ be an infinite $\cal A$-multipath.
 We know that it has an infinitely descending walk.  
Lemma~\ref{lem:stable=satisfiable} shows that it cannot be a cycle. Therefore,
it extends to infinity. 
We shall prove that, under this assumption, there is an infinitely descending thread.

The walk is made out of arcs $x[t_k,i_k]\to x[t_{k+1},{i_{k+1}}]$
for $k=0,1,\dots$ For all $t\ge t_{0}$, let $j_t$ be the
first occurrence of $t$ in the sequence; note that this is well defined.
 The walk is broken into segments leading from
$x[t,i_{j_t}]$ to $x[t+1,i_{j_{t+1}}]$, of which infinitely many are
descending.  We claim that each of this segments can be replaced with
a single arc, strict when appropriate. This implies that
SCT is satisfied.

As there is a walk from $x[t,i_{j_t}]$ to $x[t+1,i_{j_{t+1}}]$, consider
the shortest one (the shortest strict one, if appropriate).
Suppose that it consists of more than one arc.
 Then there is a
node $x[t^*,i_{j_{t^*}}]$ occurring inside the segment (not at its ends) such that
the node preceding it, $x[t^-,i_{j_{t^-}}]$ and the node suceeding it, $x[t^+,i_{j_{t^+}}]$
satisfy either $t^- , t^+ \le t^*$ or $t^- , t^+ \ge t^*$ (it may take
a little reflection to see that
such a node must exist; see Figure~\ref{fig:stable=sct}).
Then consider the two arcs---from $x[t^-,i_{j_{t^-}}]$ to $x[t^*,i_{j_{t^*}}]$
and from  $x[t^*,i_{j_{t^*}}]$ to $x[t^+,i_{j_{t^+}}]$;
they must both lie in the same MC, and
by consequence closure, there is a single arc
(strict if any of the two arcs is) to replace these two arcs.
Thus, the presumed shortest walk is not shortest.

We conclude that the shortest walk consists of a single arc, which is then
a forward arc in the MC. We conclude that an infinitely descending \emph{thread}
exists in $M$, so that SCT is satisfied.
\eprf
\begin{figure}[t]
\vspace{-0.5cm}
\begin{centering}
$\xymatrix@R=10pt@C=10pt{
 t-1 & t & t+1 & t+2 \\
& \bullet\ar@{->}[dl]\ar@{.}[u]\ar@{.}[dd] & \bullet\ar@{.}[dd]\ar@{.}[u] \\
\bullet\ar@{->}[r] & \bullet\ar@{->}[r]\ar@{.}[d] & \bullet\ar@{->}[r]\ar@{.}[d] & \bullet\ar@{->}[ul]\\
 & \bullet\ar@{->}[dr]\ar@{.}[dd] & \bullet\ar@{.}[d] & \bullet\ar@{->}[l]\\
 & & \bullet\ar@{->}[ur]\ar@{.}[d] & \\
 & \bullet\ar@/_10pt/@{->}[d]\ar@{.}[d] & \bullet\ar@{.}[d] \\
 & \bullet\ar@{->}[r]\ar@{.}[d] & \bullet\ar@/_10pt/@{->}[u]\ar@{.}[d] \\
 & & \\
}$ 

\end{centering}
\caption{Three examples of walks from $x[t,i_{j_t}]$ to $x[t+1,i_{j_{t+1}}]$:
in the first (topmost), $t^*$ may be taken as either $t-1$, or $t+2$; in the second, only as $t+2$;
in the third, as either $t$ or $t+1$.
}
\label{fig:stable=sct}
\end{figure}

\subsection{A decision algorithm.}

As an immediate corollary, we obtain a new algorithm
to decide MCS termination. Namely, 

\begin{algo} (Deciding termination of an MCS ${\cal A}$) \label{alg-sct}
\end{algo}
\be
\item Stabilize the system (e.g., by full elaboration)
\item Apply an SCT decision algorithm.
\ee
Note that since we are deciding SCT,
we can ignore any ``backward" arcs ($x_j'\to x_i$),
as well as the state invariants, in other words retain just SCT graphs.
This observation may possibly be useful in optimizing an implementation.

Another natural expectation is that it would be desirable 
in practice to
avoid full elaboration when possible, using a more economic
stabilization procedure. Such a procedure is described next. We will discuss
the efficiency of Algorithm~\ref{alg-sct} afterwards.
\subsection{A more economic stabilization}
\label{sec:stabilization}

Given an MCS $\cal B$, the following algorithm computes a stable system $\cal A$
that bisimulates it.  No variable renaming is used, so we only have to compute
the set of flow-points and the mapping $\phi$. 

\begin{algo} Stabilization by fixed-point computation \label{alg-stabilize}
\end{algo}
\be[(1)]
\item
Initialize the system $\cal A$:
flow-points in
$\cal A$ will be uniquely identified by a pair $\langle f, I\rangle$ where $f$ is the corresponding
$\cal B$ flow-point and $I$ is a state invariant. Initially,
for every $f\in F^{\cal B}$, we have $\langle f, I_f\rangle$ in $\cal A$ where
$I_f$ is the invariant from $\cal B$.
Set $\phi$ to associate $f$ with $\langle f, I_f\rangle$ and copy all 
MCs and invariants from $\cal B$.
\item
Replace all MCs with their closure under logical consequence; if an MC is unsatisfiable, delete it.
\item
Repeat the
following process until instructed to stop:
\be[(a)]
\item Search for an MC $G:\langle f, I_f\rangle \to \langle g, I_g\rangle$ which is not stable
(checking for stability is a simple graph algorithm based on
Lemma~\ref{lem:floyd}).
If no such MC is found, stop.
\item  Since $G$ is not stable, there is a relation,  $x_i \rhd x_j$,
such that
$G \vdash x_i\rhd x_j$, but $x_i \rhd x_j$ is not in $I_f$
(or a similar situation on the $g$ side).
\be[(i)]
\item Suppose that the missing relation is $x_i > x_j$.
Create two $\cal A$ flow-points to replace $\langle f, I_f\rangle$: 
$\langle f, \: I_f\land ( x_i > x_j ) \rangle$ and
$\langle f, \: I_f\land ( x_i \le x_j ) \rangle$.
If any of these points is already in $\cal A$, there is nothing more to do about it.
For a point which is new, all MC's previously leading from and to $\langle f, I_f\rangle$
have to be copied to this new flow-point. Finally the old flow-point  $\langle f, I_f\rangle$
is deleted.
\item If the  missing relation is $x_i \ge x_j$,  the flow-points created will be
$\langle f, \: I_f\land (x_i\ge x_j) \rangle$ and
$\langle f, \: I_f\land (x_i  < x_j) \rangle$. The rest is as above.
\ee
\ee
\ee

It is not hard to see that the above loop will terminate. In fact, the number of $\cal A$ flow-points
corresponding to a given $\cal B$ flow-point $f$ is bounded (by $B_n$), so at some point the set
of flow-points will stop growing, and then all size-change graphs must be stable.

Note that this algorithm will not change an SCT instance; SCT instances
are stable.

\subsection{Efficiency of analysing an MCS by reduction to SCT}

To analyse the efficiency of Algorithm~\ref{alg-sct}, we have to consider
two questions. First, what is the cost of stabilization? The cost of full
elaboration is $O(|{\cal B}| n^{2n+1})$, where $\cal B$ is the input system.
Algorithm~\ref{alg-stabilize} may take less, and it seems reasonable to
expect it to take less in many instances (where there are variables
in the abstract program that are not related to each other).
The worst-case, however, is to reach full elaboration.

Then, an SCT decision procedure has to be applied. What SCT procedure?
It is interesting to consider the
closure-based algorithm~\cite{leejonesbenamram01}, the SCT version of
Algorithm~\ref{alg-closure}; this
is apparently the most popular SCT algorithm (as witnessed by,
 e.g.,~\cite{TG:05,MV-icse06,FV:09}).
Let $\cal A$ be the stabilized system. Suppose that it has $m$ flow-points.
Its closure set can reach (at the worst case) a size of approximately
$m^2 3^{n^2}$, as there
are $3^{n^2}$ possible size-change graphs for a given pair of flow-points.
The complexity of the algorithm is thus bounded by 
$m^2 3^{n^2} \le (B_n)^2 3^{n^2}$ times some low-order polynomial,
so, succinctly, it is $2^{\Theta(n^2)}$.

Surprisingly, the worst-case complexity \emph{drops} if we assume that
the MCS has been fully elaborated (at first sight the wasteful choice).
In fact, fixing a pair of flow-points,
there are less than ${n}^{2n}$ different graphs between them
that have the downward closure property (Page~\pageref{dfn:downwardclosure}).
This property is preserved by composition, so the closure computed in deciding
SCT will be bounded by $m^2 n^{2n}$, 
where $m \le B_n = 2n^{n-1}$,
so the total complexity will be
$2^{O(n\log n)}$.
Thus, full elaboration
improves the asymptotic worst-case complexity of the closure algorithm.
However, as the next section will show, if we have a fully-elaborated
system, there is actually a polynomial-time algorithm that decides
termination (and more); this is because the SCT instance obtained is of
a special structure. Hence, there is no need to do anything as costly
as a closure computation.

\subsection{The Codish-Lagoon-Stuckey Algorithm}
\label{sec:clsAlg}

The decision algorithm suggested by~\cite{Codish-et-al:05}
is to first compute the closure set of a given MCS,
and then, to every cyclic MC, apply
``balancing'' which is similar to our stabilization but on a local basis.
After balancing, the cyclic graphs are tested, as for SCT.
Thus, the algorithm is very similar to Algorithm~\ref{alg-closure}; the LTT
is replaced by the balancing procedure followed by
Sagiv's test (which as in SCT takes a simpler form for idempotent graphs).

Clearly, this test must be equivalent to the LTT in its results; its complexity
may differ, though \emph{grosso modo} they are similar---both are low-order
polynomial in the size of the MC.  

 Codish, Lagoon and Stuckey
do not prove completeness of their analysis with respect to termination of the original
system, but to termination of each tested graph as a singleton;
however this gap is not hard to bridge. Thus, we have at our disposal 
three slightly different algorithms for deciding MCS termination (four,
when the results of the next section are taken into account); but 
it may suffice to take only two home: one closure-based algorithm (I propose
Algorithm~\ref{alg-closure}),
and one based on full-elaboration (and the continuation to be
given in the next section). The first has a higher worst-case complexity,
$m^2 2^{\Theta(n^2)}$ versus $m^2 2^{\Theta(n\log n)}$; but it has a lower
best-case complexity, which may be useful in practice. Moreover, its upper
bound drops to $m^2 2^{\Theta(n\log n)}$ for a useful class of
SCT instances, fan-in free graphs~\cite{FV:09}.

A clarification may be due regarding the assumptions made on $\val$.
Section~\ref{sec:semantics} includes the statement 
``all notions and results in this paper
work equally well with partial orders, and even partial quasi-orders.''
Full elaboration, as well as the ``economic'' stabilization, assume that the
order is total; for any $u,v\in\val$, one of $x<y$, $x=y$ and $x>y$ must hold.
However, by Lemma~\ref{lem:wf}, it is possible to extend a given partial
order to a total one. While this extension is not necessarily constructive,
this does not matter: it suffices to imagine that such an extension is being
used in order to explain the algorithm. After all, the termination condition
does not really depend on the semantic domain, since
it can be stated in pure graph terms (Section~\ref{sec:criterion}).

The results of the following section are an exception: if we ask, not only
for decision regarding termination, but for a ranking function, the order
on $\val$ must really be total,
since the ranking function descends in an order derived from it.

\section{Constructing a Global Ranking Function}
\label{sec:grf}

This section describes a ranking-function construction for 
monotonicity constraint systems. 
At the heart of the construction is an algorithm that constructs a ranking function,
if possible, for a fully-elaborated system. It does so in polynomial time,
which highlights the fact that fully-elaborated systems are a special class of SCT
instances. They are also special in that their global
ranking functions are particularly simple---a tuple of elements (constants or variables)
is associated with each flow-point, and these descend lexicographically in each and
every transition.

The rest of this section is organized as follows. The first subsection presents tools that are
used in exposing the structure of a fully-elaborated system. The second employs them to construct
ranking functions. The third puts together the result for general monotonicity constraint systems.

\subsection{Thread preservers and freezers}

In preparation for the analysis of fully elaborated systems, we define complete threads,
thread preservers and freezers and prove some essential properties.
Note: when notions of 
connectivity are applied to an MCS (for example, ``$\cal A$ is strongly connected''),
they actually concern the underlying control-flow graph.

\bdfn[complete thread]
A thread in a given multipath is 
{\em complete} if it starts at the beginning of the multipath,
and is as long as the multipath.
\edfn

\blem \label{lem:finiteM}
If a strongly connected MCS satisfies SCT, every finite multipath
includes a complete thread.
\elem

\bprf
Assume that SCT is satisfied by the given MCS; then every infinite
multipath contains an infinite thread.  Let $M$ be any finite multipath,
beginning and ending at the same flow-point.
Consider $M^\omega$, i.e., the concatenation of infinitely many copies
of $M$. From an infinite thread in $M^\omega$ one can 
clearly ``cut out'' a
complete thread in one of the copies of $M$. Thus, every finite multipath
that begins and ends at the same flow-point has a complete thread.
A multipath $M'$ that does not end at the point of departure can be
extended to a multipath that does, since we assumed strong connectivity
of the MCS. Thus, it has a complete thread.
\eprf


\bdfn[thread preserver]
Given MCS $\cal A$, a mapping
$P:F^{\cal A} \to \powerset(\{1,\dots,n\})$
is called a {\em thread preserver\/} of $\cal A$
if for every $G:f\to g$ in $\cal A$, it holds that
whenever $i\in P(f)$,
there is $j\in P(g)$ such that $x_i{\to}x_j'\in G$.
\edfn

It is easy to see that the set of thread preservers of $\cal A$
is closed under
union. Hence, there is a unique maximal thread preserver,
which we denote by $\mbox{MTP}({\cal A})$.
Given a standard representation of $\cal A$, 
$\mbox{MTP}({\cal A})$ can be found in linear time (for details see~\cite{BL:2006}).

\bdfn
A variable $x_i$ is called \emph{thread-safe} at flow-point $f$ if
every finite $\cal A$-multipath, starting at $f$,
includes a complete thread from $x_i$.
\edfn

\blem \label{lem-tp}
Let $\cal A$ be a fully-elaborated, strongly connected, terminating constraint system.
For every $f$, let $S(f)$ be the set of indices of variables that are
thread-safe at $f$. Then $S(f)$ is not empty for any $f\in F^{\cal A}$
and $S$ is a thread preserver.
\elem

\bprf
Let $M$ be any finite ${\cal A}$-multipath starting at $f$.
Observe that since
${\cal A}$ satisfies SCT and is strongly connected,
there must be a complete thread in $M$,
say starting at $x_i$.
But then $x_n$ can also start a thread (note the downward-closure
of fully-elaborated MCs). It follows that $n\in S(f)$.

We now aim to show that $S$ is a thread preserver.
Let $i\in S(f)$,
and let $G:f\to g$. Every finite multipath $M$ beginning with $G$ has
a complete thread that begins with an arc from $x_{i}$, say
$x_{i}\to x_{j_M}'$. Let $J$ be the set of all such indices $j_M$, and
$k = \max J$.
 Then $x_{i}\to x'_{k}$ is an arc of $G$, because $k\in J$;
and by the downward-closure property one can see that
every $M$ has a complete thread beginning with the arc
$x_{i}\to x_{k}'$. Hence, $k\in S(g)$
and the proof is complete.
\eprf

Informally, with a fully elaborated system, we are assured by this lemma that non-empty thread preservers
exist. The next lemma shows that given any thread preserver we can find, within it,
a \emph{singleton thread preserver}. This comes very close to identifying a ranking
function (or at least a quasi-ranking function, defined in the next subsection).

\blem \label{lem-tpmin}
Let $\cal A$ be a fully-elaborated, terminating MCS, and
$S$ a thread preserver, where $S(f)\ne \emptyset$ for all $f$.
For every $f\in F^{\cal A}$, let $i_f = \min S(f)$.
Then $P(f) = \{i_f\}$ is a thread preserver. In other words,
every MC $G:f\to g$ includes $x_{i_f}\to x_{i_g}'$.
\elem

\bprf
By the definition of a thread-preserver,
$G$ must have an arc from $x_{i_f}\to x_j'$ with ${j\in S(g)}$;
so by downward-closure, $G$ includes $x_{i_f}\to x_{i_g}'$.
\eprf

\bdfn[freezer]
Let $C:F^{\cal A}\to \{1,\dots,n\}$
denote a choice of one variable for each flow-point.
Such $C$ is called a {\em freezer\/} for $\cal A$
if for every $G\in {\cal A}$, $G\vdash x_{C(f)} = x_{C(g)}'$.
\edfn

Informally, a freezer is a singleton thread-preserver where the values are ``frozen''
since all the arcs represent equality.

\blem \label{lem-dropvar}
Let $\cal A$ be a stable MCS that satisfies SCT,
and has a freezer $C$.
If for every $f$, variable $x_{C(f)}$ is ignored,
SCT will still be satisfied.
\elem

\bprf
Let $M$ be an infinite multipath of $\cal A$; by 
the SCT property,
$M$ has an infinitely descending thread $\tau$.
Observe that $C$ induces
an infinite thread $\vartheta$ in $M$, consisting entirely of no-change
arcs.  
We claim that $\vartheta$ can have at most finitely many intersections
with $\tau$. To prove it, assume the contrary. Then there must be a
strict arc in $\tau$ between two intersections. That is, the nodes at the intersections
are connected by a descending path (via $\tau$) and by a no-change path (via $\theta$),
which is a contradiction, making this part of $M$ unsatisfiable. But, according to 
Lemma~\ref{lem:stable=satisfiable},
this cannot happen.
 It follows that $M$ has an infinitely descending thread (namely $\tau$,   minus 
some finite prefix) that avoids the frozen variables. Since this holds for any infinite multipath,
SCT is satisfied with these variables omitted.
\eprf

\subsection{Ranking functions for fully-elaborated systems}

From this point on, fix $\cal A$ to be a fully-elaborated system.

Provided that it terminates, a ranking function will be constructed. 
To precisely specify the form of this function,
we define vectors (definition from~\cite{BL:ranking2}, but simplified).

\bdfn[vectors] \label{def-Vf}
Let $n>0$ represent the number of variables in a program under consideration, and let $B>0$ be
some integer.
 $V^B$ is the set of tuples 
$\vec v = \langle v_1,v_2,\dots\rangle$ of even length,
where every even position is a variable among $\{x_1,\dots,x_n\}$, such that every
variable appears at most once;
and every odd position is an integer between 0 and $B$.
\edfn

\bdfn \label{def-value}
The value of $\vec v \in V^B$ in program state $(f,\sigma)$,
denoted $\vec v \sigma$, is obtained by
substituting the values of variables according to $\sigma$.
This results in a tuple with elements of $\val$ and integers in even and odd
positions, respectively. Such tuples are compared lexicographically.
\edfn

The functions we construct have a simple form: 
to each flow-point $f$ a vector $\vec v_f$ is associated so that
$\rho(f,\sigma) = \vec v_f \,\sigma$.  Thus, for every transition $(f,\sigma)\mapsto
 (g,\sigma')$, we shall have $\vec v_f \sigma \succ  \vec v_g \,\sigma'$, where $\succ$
 is the strict lexicographic order.

The construction is incremental. To justify the incremental
construction, we define quasi ranking functions and residual transition systems.

\bdfn
Let $\cal T$ be a transition system with state space ${\it St}$.
A \emph{quasi-ranking function}
for $\cal T$ is a function
$\rho:{\mathit St}\to W$, where $W$ is
a well-founded set,
such that $\rho(s) \ge \rho(s')$ for every $(s,s')\in {\cal T}$.

The residual transition system relative to $\rho$, denoted ${\cal T}/\rho$,
includes all (and only) the transitions of $\cal T$ which do \emph{not}
decrease $\rho$.
\edfn
%
Note that when Lemma~\ref{lem-tpmin} applies, it provides a quasi-ranking
function: $\rho(f,\sigma) = \langle x_{i_f}\rangle \sigma$.

\newcommand{\cat}{\mathop{++}}


The next couple of lemmas are
quite trivial but we spell them out because they clarify
how a ranking function may be constructed incrementally.
We consider the codomain of all our functions to consist of
lexicographically-ordered tuples over ``scalars'' (a scalar is
either a constant or a variable) and
we use the notation $v\cat u$ for concatenation of tuples.

\blem
Assume that $\rho$ is a quasi-ranking function for $\cal T$,
and $\rho'$ a ranking function for ${\cal T}/\rho$; then $\rho\cat \rho'$
is a ranking function for $\cal T$.
\elem

\blem
Assume that the CFG of ${\cal A}$ consists of a set $C_1,\dots,C_k$
of mutually disconnected components (that is, there is no arc 
from $C_i$ to $C_j$ with $i\ne j$). If for every $i$,
$\rho_i$ is
a ranking function for ${\cal A}$ restricted to $C_i$,
then
$\cup_i \rho_i$ is a ranking function for ${\cal A}$.
\elem

\blem
Suppose that the CFG of ${\cal A}$ consists of several strongly connected
components (SCCs). Let $C_1,\dots,C_k$ be a reverse topological ordering
of the components. Define a function $\rho(s)$ for $s = (f,\sigma)$
as the index $i$ of the component $C_i$ including $f$. Then $\rho$ is
a quasi-ranking function (with co-domain $[1,k]$) and it is strictly
decreasing on every transition represented by an inter-component arc.
\elem

The following algorithm puts all of this together.  Note: a CFG
whose arc set is empty is called \emph{vacant}. A strongly connected
component whose arc set is empty is called \emph{trivial} (it may
have connections to other components).

\begin{algo} (ranking function construction for ${\cal A}$) \label{alg-A}
\end{algo}
The algorithm assumes that $\cal A$ is fully elaborated. If $\cal A$ terminates,
a ranking function will be returned. Otherwise, the algorithm will fail.

We assume that the representation of $\cal A$ allows for ``hiding'' certain 
variables of any given flow point. This affects
subsequent MTP computations, which will ignore the hidden variables.
\be[(1)] 
\item
List the SCCs of ${\cal A}$ in reverse-topological order. For each $f\in F^{\cal A}$,
let $\kappa_f$ be the
position of the SCC of $f$. Form ${\cal A}'$ by deleting
all the inter-component transitions. If ${\cal A}'$ is vacant,
return $\rho$ where $\rho(f,\sigma) = \kappa_f$.
\item
For each SCC $C$,
compute the MTP, using the algorithm in~\cite{BL:2006}.
If empty, report failure and exit.
\item \label{step:addvar}
   For every $f$, let $x_{i_f}$ be the lowest
   MTP variable of $f$.
\item \label{step:nonstrict}
     For every graph $G:f\to g$, if it
     includes $x_{i_f}\strict x_{i_g}'$,
     delete the graph from ${\cal A}'$;
     otherwise, retain the graph but hide $x_{i_f}$.
\item
 For every $f$,  let $\rho(f,\sigma) = \langle \kappa_f, x_{i_f}\rangle\sigma$.
\item
If ${\cal A}'$ is now vacant, return $\rho$.
Otherwise,
compute a ranking function $\rho'$ recursively for ${\cal A}'$,
and return $\rho\cat \rho'$. 
\ee

\subsubsection*{Correctness.}
We claim that the abstract program ${\cal A}'$, passed to the recursive call, always
represents the residual transition system ${\cal T}_{\cal A}/\rho$. This should be clear when
we delete graphs that strictly decrease $\rho$. The less-trivial point
is the treatment of graphs $G$ where the MTP arc $x_{i_f} \to x_{i_g}$
is non-strict (Step~\ref{step:nonstrict}).
To obtain the
residual system precisely we should have replaced the inequality constraints
with equalities:
$x_{i_f} = x_{i_g}'$. However, having done so, the set of indices
$C(f) = i_f$
becomes a freezer, and therefore can be ignored (Lemma~\ref{lem-dropvar}).

Hiding the ``frozen'' variables ensures that these variables will not
be used again in $\rho'$.  So in the final tuple
$\rho(f,\sigma)\cat\rho'(f,\sigma)$, each variable will occur at most once. 
This shows, in particular, that the recursion always terminates, which means that the residual
transition system is eventually vacant, and when this happens, we have a ranking function.

\subsubsection*{Complexity.}
The algorithm will make at most $n$ passes in which two elements are added to the 
tuple. The costly part of a pass is the MTP computation which takes time linear in the sum of
sizes of all the MC graphs, that is, $|{\cal A}| n^2$  (recall that $|{\cal A}|$ is the number of MCs).
In subsequent passes, the system is diminished, so an upper bound on the total running time
is $O(|{\cal A}| n^3)$.

\subsection{Ranking functions for all!}

So far, we have constructed ranking functions for fully-elaborated systems.
To construct a ranking function for a general MCS $\cal B$,
we first transform it into a fully-elaborated
$\cal A$ using Algorithm~\ref{alg-fe}.   Then, Algorithm~\ref{alg-A} can be applied.
The ranking function for $\cal A$ can be translated to one for $\cal B$, as the next lemma shows.

\blem \label{lem-getrf}
If $\cal A$ simulates $\cal B$ \emph{deterministically}, any ranking function
for $\cal A$ can be transformed into a ranking function for $\cal B$.
\elem

\bprf
Let $\rho$ be the $\cal A$ ranking function. The $\cal B$ ranking function is
defined for state
 $(f,\sigma)$ as $\rho(g,\sigma')$ where $g$ is the unique point
 such that $(f,g)\in\phi$  and
 $\sigma' \models I_{g}$,
 where $\sigma' = \sigma\circ(\psi_{g,f})$.
\eprf

When $\cal A$ is fully elaborated, we have a function $\rho$ where 
$\rho(g,\sigma)$ is a fixed vector $\vec v_g$ for every $g\in F^{\cal A}$;
thus the resulting $\cal B$ ranking function has the form 

\[
\rho(f,\sigma) = \left\{\begin{array}{ccc}
\vec v_{g_1}(\sigma\circ\psi_{g,f})	&\quad\mbox{if}\quad&	\sigma\circ\psi_{g,f}\models I_{g_1}\\
\vec v_{g_2}(\sigma\circ\psi_{g,f})	&\quad\mbox{if}\quad&	\sigma\circ\psi_{g,f}\models I_{g_2}\\
\dots\\
\vec v_{g_t}(\sigma\circ\psi_{g,f})	&\quad\mbox{if}\quad&	\sigma\circ\psi_{g,f}\models I_{g_t} \\
\end{array}\right.
\]
where $g_1,\dots,g_t$ are the $\cal A$ flow-points associated with $\cal B$
flow-point $f$.

This function may be simplified by combining rows that have the same vector as the function's value,
so we do not have to list $B_n$ rows. 
One example of such a function appears in the introduction (conclusion of Example~\ref{ex:cls}).
For another example,
consider the MCs depicted in Figure~\ref{fig-ex-m03}.
A ranking function of the form we consider is
\[ \rho(x_1,x_2,x_3,x_4) = \left\{\begin{array}{cl}
        \langle 1,x_1,1,x_3\rangle & \mbox{if $x_1> x_2$} \\
        \langle 1,x_2,1,x_4\rangle & \mbox{if $x_1 \le x_2$}. 
\end{array}\right.
\]
(obviously, the 1-valued entries can be dropped). 

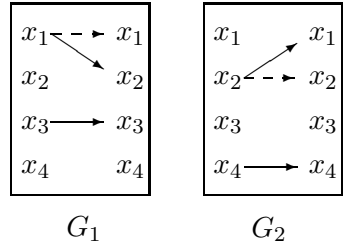
\begin{fig0}{MCs for ranking function example.
There is a single flow-point.}{t}
{fig-ex-m03}
{\setlength{\unitlength}{0.4pt}
\begin{centering}
\begin{picture}(320,200)(180,40)

	      \put(188,85){\framebox(130,180){}}
	      \put(197,231){$x_1$}
	      \put(197,189){$x_2$}
	      \put(197,147){$x_3$}
	      \put(197,105){$x_4$}
	      \put(225,237){\vector(3,-2){50}}
	      \put(225,153){\vector(1,0){50}}

 		  \dashline[20]{10}(225,237)(274,237)
 		  \put(270,237){\vector(1,0){5}}
	      \put(287,231){$x_1$}
	      \put(287,189){$x_2$}
	      \put(287,147){$x_3$}
	      \put(287,105){$x_4$}
            \put(240,45){$G_1$}
	      
	      \put(370,85){\framebox(130,180){}}
	      \put(379,231){$x_1$}
	      \put(379,189){$x_2$}
           \put(379,147){$x_3$}
           \put(379,105){$x_4$}
	      \put(408,195){\vector(3,2){50}}
           \put(408,111){\vector(1,0){50}}

		 \dashline[20]{10}(408,195)(453,195)
 		  \put(450,195){\vector(1,0){5}}
	      \put(469,231){$x_1$}
	      \put(469,189){$x_2$}
           \put(469,147){$x_3$}
           \put(469,105){$x_4$}
	      \put(415,45){$G_2$}
\end{picture}

\end{centering}} 
\end{fig0}

\bthm
Let ${\cal B}$ be a
terminating MCS, with $m$ flow-points and $n$ variables per point.
There is a ranking function $\rho$ for $\cal B$ where
$\rho(f,\sigma)$ is defined by a set of elements of $V^{mB_n}$, each one associated with
certain inequalities on variables, which define the region where the given vector is the function value.
There are at most $B_n$ different vectors for any flow-point. 
The complexity of constructing $\rho$ is 
$O(|{\cal B}|\cdot n^{2n+1})$.
\ethm

\bprf
Fully-elaborating $\cal B$ takes $O(|{\cal B}| n^{2n+1})$ time and creates a system
with $O(|{\cal B}| n^{2n-2})$ MCs.  The running time of Algorithm~\ref{alg-A} on the
result is $O(|{\cal A}| n^3) = O(|{\cal B}| n^{2n+1})$.  
\eprf

\subsubsection*{Remarks.} 
This is the first algorithm to construct explicit ranking functions for any terminating MCS,
but even when restricting attention to SCT instances, the results improve
upon previous publications. The improvement over~\cite{BL:ranking2}
is that any positive instance can be handled;
the improvement over~\cite{Lee:ranking} is that in that work, the vectors
were possibly doubly exponential in length (as a function of $n$)
and the complexity of the construction was
only bounded by a triply exponential function.

The complexity of our construction is optimal (or very nearly so) in
two senses. 
\be[$\bullet$]
\item
\cite{BL:ranking2} considers ranking functions of the form described above, and
shows that there are SCT instances with $n$ variables 
that require $(n-1)!$ distinct vectors per flow-point.
Our construction yields the close upper bound $B_n$.
\item
In a vector, each variable appears at most once. This is clearly optimal since they may all be
involved in the termination proof. This also means that the codomain of the ranking functions 
has the smallest necessary ordinal (in general).
\ee

Finally let us remark that since the processing of a fully-elaborated system is polynomial-time,
it is advantageous to use it rather than applying any general SCT decision
procedure, even if all we want is a yes-no answer.

\section{Rooted Versus Uniform Termination}
\label{sec:rooted}

In this paper, the notion of termination used was 
{\em uniform termination}, which means that there must be no cycles in the 
whole state space of the modelled transition system. 
Practically, it may be desireable
to account for
\emph{rooted termination}, when only computation paths beginning at a given
initial point $f_0$ (and satisfying its state invariant) are considered, thus avoiding
some false alarms. 

In this paper, uniform termination was chosen in favor of simplicity. However,
   It is not be difficult to adapt our methods to rooted termination;
\cite{leejonesbenamram01} shows how to do that for SCT.

\section{Conclusion and Research Questions}

We studied the MCS abstraction, an appealing extension of the Size-Change
Termination framework, showing how several elements of the theory of SCT can
be achieved in the stronger framework: sound and complete termination criteria,
closure-based algorithms based on local termination tests, and 
the construction of glocal ranking functions
(which improves on previously-published work even for SCT).
A key technique was refining the abstraction, using state invariants to separate out
the behaviour under different assumptions on the relative order of variables. We showed
that MCS termination can be simplified in this way not only to SCT termination, but to an easy
case thereof.

The contribution of this paper is theoretical;
hopefully, it will trigger further research, moving
towards the practical application of the theory.

The algorithms in this article were aimed at simplicity of presentation and
analysis and can certainly be improved.
If the purpose is just to decide termination (and possibly produce a 
non-terminating multipath if one exists),
there is a choice between a closure-based algorithm
(Section~\ref{sec:closure}) and an algorithm based on full elaboration
(Section~\ref{sec:grf}).
  While theoretically, full elaboration reduces
the exponent in the worst-case time from $\Theta(n^2)$ to $\Theta(n\log n)$,
avoiding it may well
be more efficient in practice. The intriguing effectiveness of the 
(theoretically worst) closure-based algorithms was discussed,
in the context of SCT,
 by Fogarty and Vardi~\cite{FV:09}.
 
Algorithm~\ref{alg-A} has been implemented in Java, but so far has only been tried out
with toy examples, so it is too early to speak of an empirical evaluation.
As expected, memory fills up quickly when the number of variables is enlarged.
It seems clear that in 
a practical implementation, both for deciding termination and for constructing
ranking functions, avoiding unnecessary combinatorial explosion is imperative.
Some tactics that should probably be used include an initial analysis to identify the
subset of variables that are pertinent to termination~\cite{Albert.et.al-sac08,MT:09},
and---obviously---handling one SCC of the original system at a time, instead of fully-elaborating
all at once.
Furthermore, we
may choose to resort to heuristics that sacrifice completeness for efficiency.
In the context
of SCT analysis, some heuristics have been studied
\cite{CLSS:esop2006,BL:2006,BC:08:TACAS}, 
and similar approaches can conceivably be useful with MCS.

A possible extension of this work is
modeling arbitrary Boolean combinations of order constraints
(including disjunctions)
in a direct way (rather than by converting to disjunctive normal form,
as discussed in Section~\ref{sec:mcs}).
It may be interesting to find out if 
this has practical interest.

Finally, 
perhaps the most appealing aspect of MCS, compared to the SCT abstraction, is
its usefulness in the integer domain. Adapting the theory of monotonicity constraint
termination to the integer domain is to be the subject of a forthcoming paper.


\subsubsection*{Acknowledgments.} The author thanks
Chin Soon Lee for inspiration; the LIPN laboratory at Universit\'e
Paris 13, where part of this work was done, for hospitality; Ariel
Snir for comments on the manuscript (and for programming); the editor,
Neil Jones, who pointed out an error in a previous version, and the
anonymous referees, who have provided exceptionally thorough reviews
and many suggestions for improvement.

\bibliographystyle{plain}
\bibliography{sct}
\vspace{-40 pt}

\end{document}